\documentclass{aa}  
\bibpunct{(}{)}{;}{a}{}{,} 
\usepackage{graphicx}
\usepackage{lscape}
\usepackage{txfonts}
\usepackage{xcolor}

\begin{document}

   \title{The RGB tip of galactic globular clusters and the revision of the bound of the axion-electron coupling.}

   \author{O. Straniero\inst{1,2} 
         \and
       C. Pallanca\inst{3,4}
          \and 
        E. DAlessandro\inst{4}
            \and
        I. Dom\'inguez\inst{5}
           \and
        F. R. Ferraro\inst{3,4}   
          \and
        M. Giannotti\inst{6} 
          \and
        A. Mirizzi\inst{7,8}  
          \and
        L. Piersanti\inst{1}
        }

   \institute{INAF, Osservatorio d'Abruzzo,
              Via M. Maggini, I-64100 Teramo, Italy
                \and
             INFN, Laboratori Nazionali del Gran Sasso, Italy 
                \and
             Dipartimento di Fisica e Astronomia, Universit\`a di Bologna, Via Gobetti 93/2, I-40129 Bologna, Italy 
               \and
             INAF, Osservatorio di Astrofisica e Scienza dello Spazio di Bologna, Via Gobetti 93/3, I-40129 Bologna, Italy 
                \and
             Departamento de Fis\'ica Teor\'ica y del Cosmos, Universidad de Granada, ES-18071 Granada, Spain 
                \and
             Physical Sciences, Barry University, 11300 NE 2nd Ave., Miami Shores, FL-33161, USA 
             \and
             Dipartemento Interateneo di Fisica ``Michelangelo Merlin'', Via Amendola 173, I-70126 Bari, Italy
                \and 
             INFN, Sezione di Bari, Via Orabona 4, 70126 Bari, Italy
             }

   \date{}

 
  \abstract
   {The production of neutrinos by plasma oscillations is the most important energy sink process operating in the degenerate core of low-mass red giant stars. This process counterbalances the release of energy induced by nuclear reactions and gravitational contraction, and determines the luminosity attained by a star at the moment of the He ignition. This occurrence coincides with the tip of the red giant branch (RGB), whose luminosity is extensively used as calibrated standard candle in several cosmological studies.}
   {We investigate the possible activation of additional energy sink mechanisms, as predicted by many extensions of the so-called {\it Standard Model}. In particular,  we test the possible production of axions or axion-like particles, mainly through their coupling with electrons.
   }
   {By combining Hubble Space Telescope  (HST) and ground based optical and near-infrared photometric samples, we derive the RGB tip absolute magnitude of 22 galactic globular clusters (GGCs). The effects of varying the distance and the metallicity scales are also investigated. Then we compare the observed tip luminosities with those predicted by state-of-the-art stellar models that include the energy-loss due to the axion production in the degenerate core of red giant stars.}
   {We find that theoretical predictions including only the energy-loss by plasma neutrinos are, in general, in good agreement with the observed tip bolometric magnitudes, even though the latter are $\sim 0.04$ mag brighter, on the average. This small shift may be the result  of systematic errors affecting the evaluation of the RGB tip bolometric magnitudes or, alternatively, it could be ascribed to an axion-electron coupling causing a non-negligible thermal production of axions. In order to estimate the strength of this possible axion sink, we perform a cumulative likelihood analysis using the RGB tips of the whole set of 22 GGCs. All the possible source of uncertainties affecting both the measured bolometric magnitudes and the corresponding theoretical predictions are carefully considered. As a result, we find that the value of the axion-electron coupling parameter that maximizes the likelihood probability is $g_{ae}/10^{-13}\sim 0.60^{+0.32}_{-0.58}$. This hint is valid, however, if the dominant energy sinks operating in the core of red giant stars are standard neutrinos and axions coupled with electrons. Any additional energy-loss process, not included in the stellar models, would reduce such a hint. Nevertheless, we find that values $g_{ae}/10^{-13}>1.48$ can be excluded with a 95\% of confidence.}
  {The new bound we find represents the most stringent constraint for the axion-electron coupling available so far.  The new scenario that emerges after this work represents a greater challenge for future experimental axion search. In particular, we can exclude that the recent signal seen by the XENON1T experiment was due to solar axions.}


   \maketitle
   
 
\section{Introduction}\label{sec:intro}
The observed properties of stars compared to model predictions have been often used to constrain various physical processes occurring  in stellar interiors. In particular, the brightest stars on the RGB of a globular cluster may probe some processes responsible for the energy generation, such as nuclear reactions,  as well as the efficiency of energy sinks, the most important of which is the plasma neutrino emission.  Indeed, in the core of  these low-mass stars, which are close to the He ignition, the temperature attains $\sim 100$ MK, while the central density is $\sim 10^6$ g cm$^{-3}$. According to the Standard Model, in this condition free electrons are highly degenerate and neutrinos are efficiently produced by plasma oscillations. Once produced, neutrinos escape from the stellar core, a process that subtracts the thermal energy spent for their production. On the other hand, degenerate electrons are excellent heat conductors and, in turn, they allow an efficient redistribution of the thermal energy. In addition, the high pressure of degenerate electrons hampers the core contraction and the consequent release of gravitational energy. As a result of the combined effects of neutrino production and electron degeneracy, the maximum temperature moves outward, to a layer placed approximately at 0.2 M$_\odot$ from the center, and the He ignition is delayed. In this framework, the luminosity of the RGB tip, which is the luminosity of a star at the off-center He ignition,  is determined by the efficiency of these physical processes.  

Although the scenario we have described so far is widely accepted, additional energy sink processes, as predicted by the most popular extensions of the  Standard Model, may produce sizeable modifications of our understanding of the RGB evolution and, in particular, they may affect the theoretical predictions of the RGB tip luminosity. Since the RGB tip luminosity is also considered a reliable standard candle, often used in cosmological studies \citep[see, e.g.,][]{freedman2020},
it is important to investigate the possible impact of these additional energy-loss processes. Among them, the most appealing is probably the production of axions or, more generally, of axion-like particles (ALPs). The existence of axions, small and weak interactive pseudo-scalar particles, has been firstly proposed more than 40 yr ago to solve the so-called strong CP problem, i.e., the lack of evidences of a violation of the charge-parity symmetry by strong interactions \citep{PQ1977,weinberg1978,wilczek1978}.  ALPs also emerge as pseudo Nambu-Goldstone bosons of spontaneously broken global symmetries in high-energy extensions of the Standard Model.   

Like plasma-neutrinos, axions or ALPs may be produced in the core of a RGB star by thermal processes, thus modifying the internal energy budget.  
The general rule is well known \citep{raffelt1994}: the larger the production rate of weak interactive particles 
produced by some thermal process, the brighter the tip of the RGB. For this reason, the RGB tip luminosity can be used to constrain the properties of neutrinos and other WISPs (weak interactive slim particles).   
In case of axions, Bremsstrahlung should be the most efficient production mechanism in RGB star, while other processes, like Primakoff and Compton, are suppressed because of the
high electron degeneracy.\footnote{In a recent paper \citep{straniero2019} we reviewed the axion production processes in stellar interiors and we provided the corresponding energy-loss rate to be used in stellar model calculations.} 

In a previous attempt, \citet{viaux2013} made use 
of  I-band photometric data of NGC 5904 (M5), a well studied globular cluster of the Milky Way, to derive an upper bound for the strength of the axion-electron coupling (g$_{ae}$). They found 
g$_{ae}<4.3\times 10^{-13}$ (95\% C.L.). 
In this paper we will try to improve this constraint. 
Two are the major sources of  uncertainty in the evaluation of the RGB tip luminosity, namely,  the bolometric corrections and the cluster distance. In addition to that, since the evolutionary timescale is rather short for stars in the brightest portion of the RGB, only few of them are usually observed within 2 or 3 magnitudes from the RGB tip. This occurrence hampers the clear identification of the RGB tip. In order to reduce this uncertainties, 
we will extend the M5 constraint to other clusters. First of all, we will consider V and I photometries of other two well studied clusters, NCG 362 and NGC 104 (47 Tuc). In order to collect  large and more complete samples of RGB stars, we will combine catalogs of high spatial resolution HST images of the most crowded central regions with ground based observations of the external portions of the cluster fields.  In addition, NGC 362 and NGC 104 are the only two GGCs for which a reliable distance can be estimated from the parallaxes reported in the latest Gaia Data Release 2 \citep{chen2018}. Then, we will extend our analysis to a  sample of 22 GCs, for which near-infrared (near-IR) photometries are available after a wide observational campaign performed by \citet{ferraro2000, sollima2004, valenti2004a, valenti2004c}. The availability of JHK observations present several advantages. Indeed, 
the brightest RGB stars are rather cool objects and their 
spectral energy distribution is dominated by near-IR light. 
For this reason, near-IR bolometric corrections of RGB stars are
more reliable than that required at shorter wavelengths, allowing a more 
reliable derivation of the bolometric magnitudes \citep[see. e.g.,][]{buzzoni2010}. In addition, some of the brightest RGB stars are long period variables, whose amplitude may be quite large in the optical spectral range while it is much smaller in the near-IR \citep{kiss2003}.  

The models used to estimate the RGB tip luminosity of GGCs are presented in section \ref{sec:methods}, while the adopted photometric samples and the derivation of the absolute bolometric magnitude of the RGB tip are described in section \ref{sec:clusters}. The statistical analysis and its results are discussed in section \ref{sec:analysis}. Summary and conclusions follow.

\begin{figure}
   \centering
   \includegraphics[width=9.0cm]{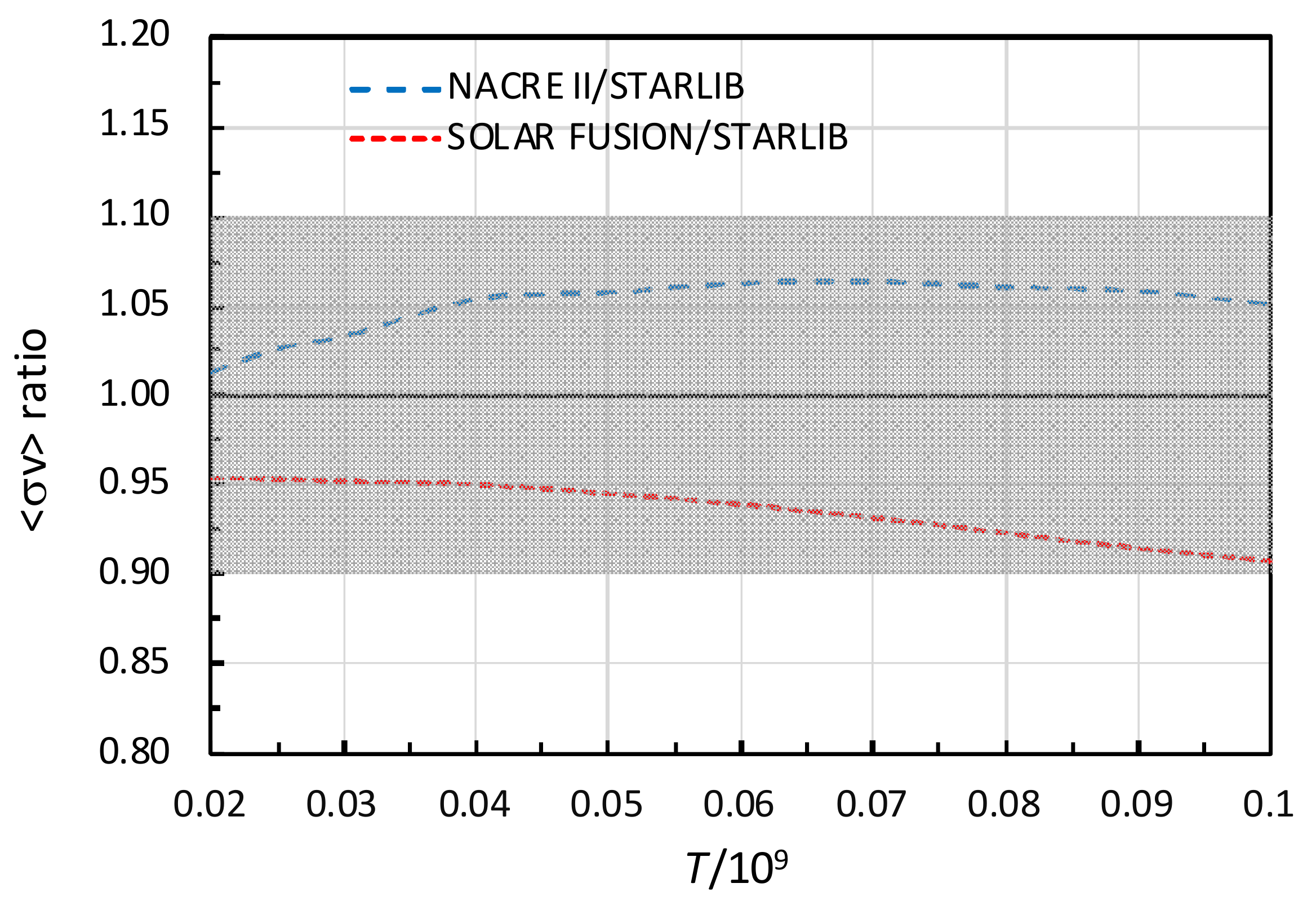}
   \caption{$^{14}$N$(p,\gamma)^{15}$O reaction rates suggested by the collaborations NACRE II and Solar Fusion are compared to the rate reported in the STARLIB repository. The shaded area represents the assumed uncertainty.}
   \label{n14p}
\end{figure}

  \begin{figure}
   \centering
   \includegraphics[width=9.0cm]{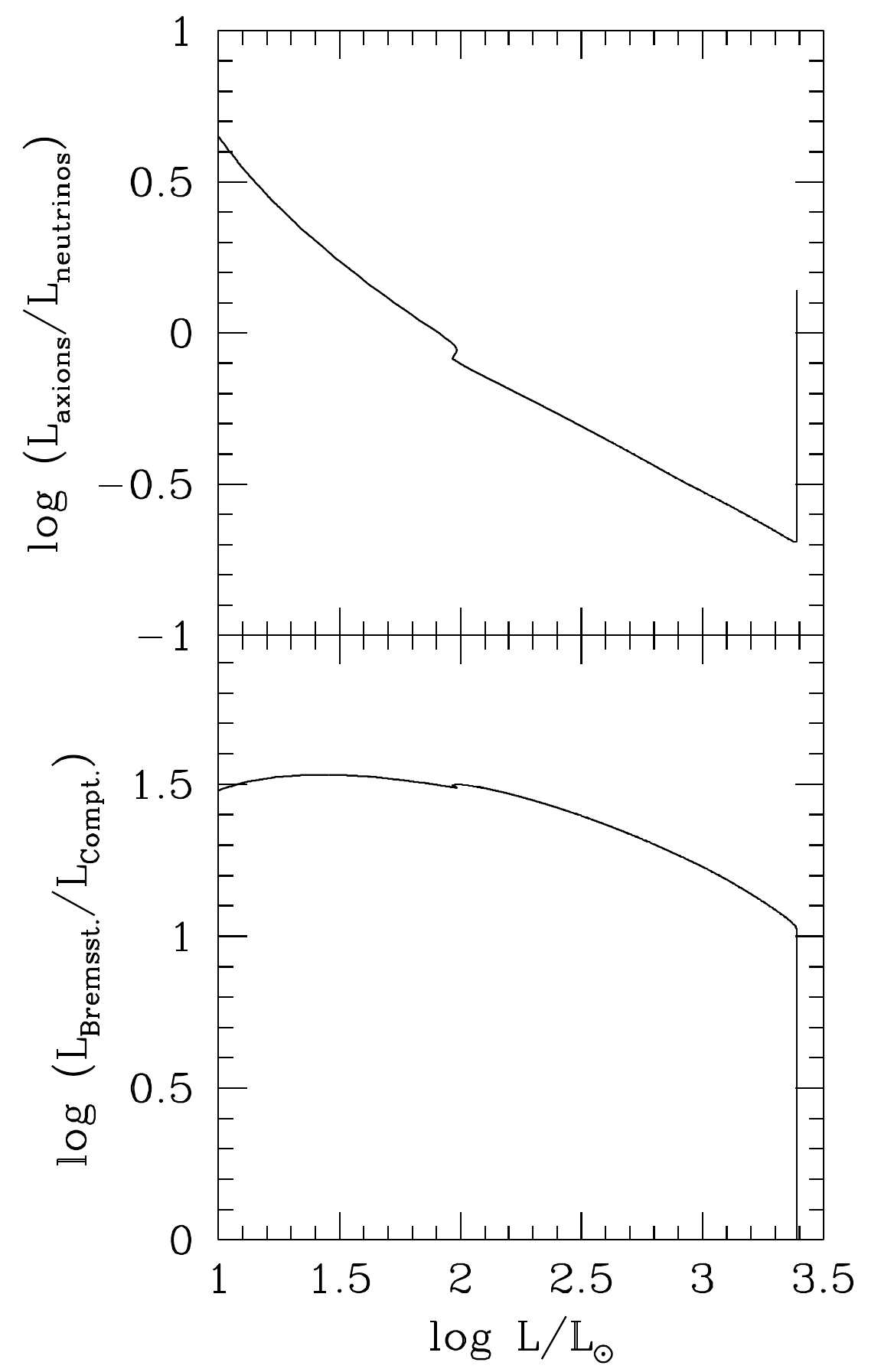}
   \caption{Upper panel: evolution of the ratio between axion and neutrino luminosites, from the base of the RGB ($\log L/L_\odot=1)$ to the RGB tip, for a model with  $M=0.82$ $M_\odot$, $Y=0.25$, $Z=0.001$. Only the coupling with electrons has been switched on ($g_{13}=1$). Lower panel: evolution of the ratio between Bremsstrahlung and Compton axion luminosities.}
   \label{loss}
\end{figure}

  \begin{figure}
   \centering
   \includegraphics[width=9.0cm]{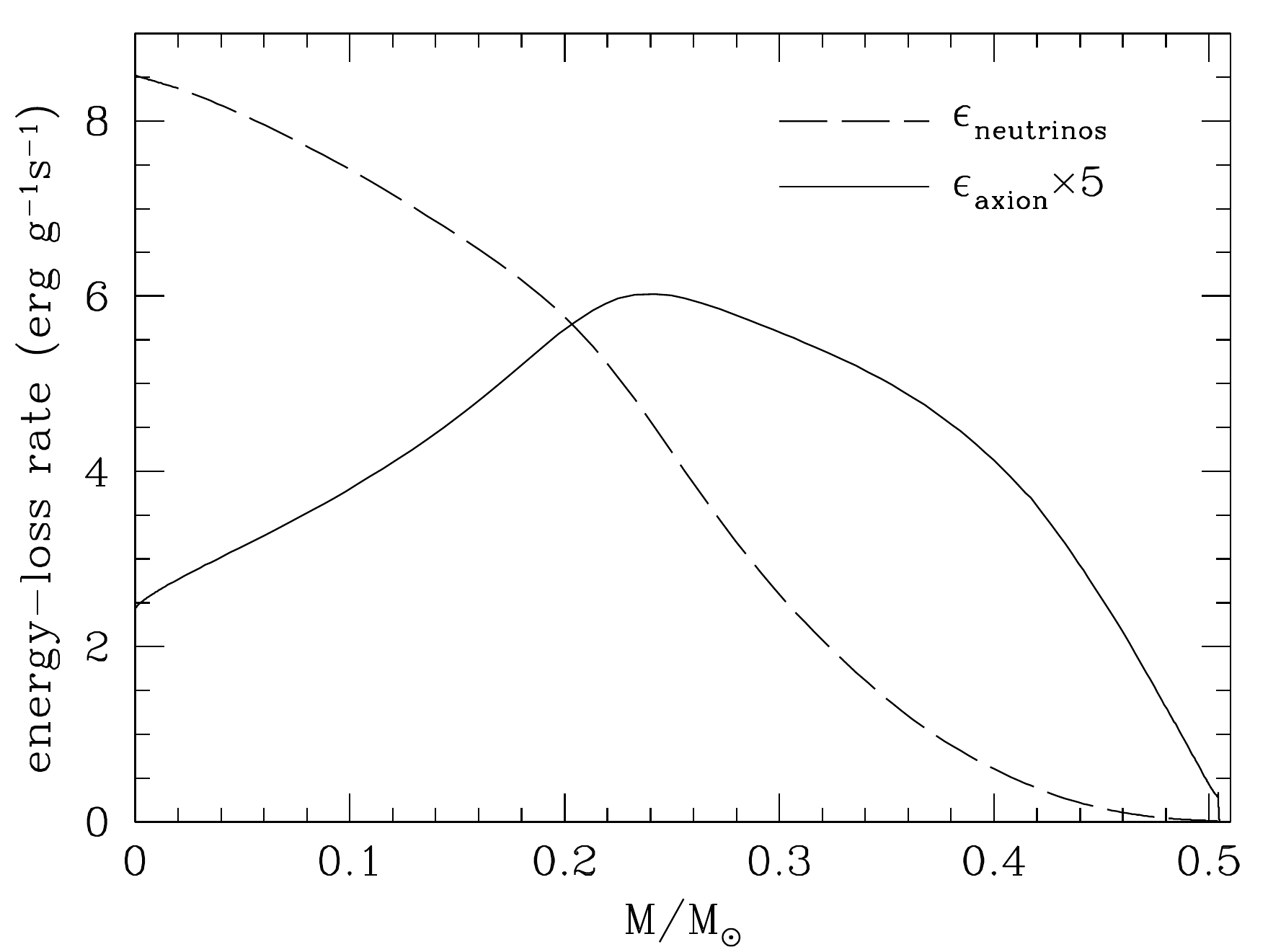}
   \caption{Neutrinos and axions energy-loss rate within the core of a RGB model close to the RGB tip (just before the He ignition). Model parameters as in figure \ref{loss}}
   \label{flash}
\end{figure}

\begin{table}
	\centering
	\caption{Theoretical predictions of the tip bolometric magnitudes for globular cluster models with $age=13$ Gyr and $Y=0.25$. From column 1 to 5: metallicity (mass fraction), $[M/H]=\log (Z/X) - \log (Z/X)_\odot$, coupling parameter $g_{13}=g_{ae}\times 10^{13}$, tip bolometric magnitude $M_{bol}=4.75-2.5\log L/L_\odot$, difference with the tip bolometric magnitude of the $g_{13}=0$ model.}
	\label{tab:modelli}
	\begin{tabular}{ccccc} 
		\hline 
Z  &   [M/H] &  $g_{13}$ & $M_{bol}$  & $\delta M_{bol}$ \\
\hline
0.0001 & -2.156 & 0.0 & -3.426 & 0.000 \\
0.0001 & -2.156 & 0.5 & -3.455 & 0.029 \\
0.0001 & -2.156 & 1.0 & -3.510 & 0.084 \\
0.0001 & -2.156 & 1.5 & -3.591 & 0.165 \\
0.0001 & -2.156 & 2.0 & -3.699 & 0.273 \\
0.0001 & -2.156 & 2.5 & -3.799 & 0.373 \\
0.0001 & -2.156 & 3.0 & -3.901 & 0.474 \\
0.0001 & -2.156 & 4.0 & -4.082 & 0.656 \\
0.0001 & -2.156 & 6.0 & -4.365 & 0.939 \\
  \hline 
0.001  & -1.156 & 0.0 & -3.643 & 0.000 \\
0.001  & -1.156 & 0.5 & -3.667 & 0.024 \\
0.001  & -1.156 & 1.0 & -3.715 & 0.072 \\
0.001  & -1.156 & 1.5 & -3.791 & 0.148 \\
0.001  & -1.156 & 2.0 & -3.881 & 0.237 \\
0.001  & -1.156 & 2.5 & -3.972 & 0.328 \\
0.001  & -1.156 & 3.0 & -4.062 & 0.419 \\
0.001  & -1.156 & 4.0 & -4.228 & 0.585 \\
0.001  & -1.156 & 6.0 & -4.513 & 0.870 \\
 \hline                                      
0.006  & -0.374 & 0.0 & -3.793 & 0.000 \\
0.006  & -0.374 & 0.5 & -3.815 & 0.022 \\
0.006  & -0.374 & 1.0 & -3.855 & 0.062 \\
0.006  & -0.374 & 1.5 & -3.915 & 0.123 \\
0.006  & -0.374 & 2.0 & -3.995 & 0.202 \\
0.006  & -0.374 & 2.5 & -4.074 & 0.281 \\
0.006  & -0.374 & 3.0 & -4.157 & 0.364 \\
0.006  & -0.374 & 4.0 & -4.315 & 0.522 \\
0.006  & -0.374 & 6.0 & -4.581 & 0.788 \\
\hline

\end{tabular}
\end{table}


 \section{Theoretical recipes}\label{sec:methods}
 \subsection{The RGB tip luminosity from  stellar models} \label{sec:theory}
 The stellar models we use in the following analysis have been computed by means of the Full Network Stellar evolution code (FuNS). This code has been extensively used to calculate models of stars of any mass and chemical composition and the related nucleosynthesis. Recently, we have included in the energy conservation equation the terms that account for the energy loss caused by an axion production possibly induced by various thermal processes, in particular, Primakoff, Bremsstrahlung, $e^+e^-$ pair annihilation and Compton's scattering on electrons or nuclei. The current version of the FuNS code and the relevant input physics are described in \citet{straniero2019} (see, in particular, section 2 and the appendix). It also  contains a detailed description of the algorithms we use to calculate the axion energy-loss rates\footnote{The same algorithms are also valid for the production of more general ALPs. Therefore, the conclusions of the present work apply to axions as well as to ALPs.}. Let us summarize here the most important input physics adopted in the present work.
 
 In general, the convective boundaries are fixed according to the Ledoux criterion.
 No convective overshoot is applied. In the convective zones, the temperature gradient is computed according to the mixing-length theory, as described in \citet{cox&giuli}. The mixing length parameter, i.e. $\alpha=\Lambda/H_P=1.82$, has been calibrated so as to reproduce the solar radius \citep{piersanti2007}. 
The FuNS code may also account for stellar rotation \citep{piersanti2013} and microscopic diffusion \citep{straniero1997,piersanti2007}, but these processes are not included in the present stellar models. 

 The FuNS has been optimized to handle large nuclear networks, as those required 
to follow the neutron-capture nucleosynthesis in AGB stars \citep[see][]{straniero2006}. 
For the purpose of the present work, we have considered 19 reactions for the H burning, those describing a full pp-chain and CNO-cycle, and an $\alpha$-chain for the He burning. All the stable isotopes of H, He, Li, Be, C, N, O plus $^{20}$Ne are explicitly included in the nuclear network.
 The reaction rates are from the version 6 of the STARLIB database \citep{starlib}.
 Note that most of the H and He burning reaction rates in this database are based on available experimental data.
 Rates of reactions for which no experimental information exists, or extrapolation  to low/high temperatures are required, have been obtained by means of  statistical  (Hauser–Feshbach)  models  of  nuclear reactions \citep[TALYS code,][]{talys2008}.
 The luminosity of the RGB tip is particularly sensitive to the adopted rates of the $^{14}$N$(p,\gamma)^{15}$O and the triple-$\alpha$ reactions. The first reaction is the bottleneck of the CNO cycle and, in turn, it determines the rate of growth of the core mass during the RGB phase. This reaction has been the object of an intense experimental campaign performed by the LUNA collaboration down to a center of mass energy of 70 keV, which corresponds to the Gamov's peak for the typical temperature of the shell H-burning in evolved RGB stars \citep{formicola2004,imbriani2004,lemut2006}. Our knowledge of this reaction is illustrated in figure \ref{n14p}, where the rate reported in the STARLIB repository \citep{starlib}, is compared to those reported in \citet{adelberger2011} and  in the NACREII repository \citep{nacreII}. Note that in the temperature range 20 to 100 MK, the differences are always smaller than 10\%.  
 
 On the other hand, the triple-$\alpha$ reaction determines the He ignition and, in turn, it fixes the brightest point attained by a RGB star. The most recent experimental investigation of this reaction has been reported by \citet{fymbo2005}. Nevertheless, the widely adopted reaction rates, such as those reported in STARLIB or REACLIB \citep{reaclib}, are all based on the NACRE rate \citep{angulo1999}. In any case, in the temperature range 100 to 300 MK, the differences between the\citet{fymbo2005} and the NACRE rates are less than the quoted $20\%$ error.  
 Finally, electron screening corrections are computed according to \citet{dewitt1973,graboske1973,itoh1979}, for weak, intermediate and strong ion coupling, respectively.
 Other FuNS input physics are here listed:

\begin{itemize}
  \item Neutrino Energy Loss: 
     \begin{itemize}
        \item Plasma $-$ \citet{haft1994}
        \item Photo $-$  \citet{itoh1996a}  
        \item Pair  $-$  \citet{itoh1996a}  
        \item Bremsstrahlung $-$ \citet{dicus1976}
        \item Recombination $-$ \citet{BPS1967}
     \end{itemize} 
  \item Radiative Opacity:    
      \begin{itemize}
         \item \citet{AF1994}+\citet{OPAL}+\citet{LAOL}
      \end{itemize}       
  \item Electron Conductivity: 
      \begin{itemize}
        \item \citet{potekhin1999a,potekhin1999b}
      \end{itemize}
  \item Equation of State:
     \begin{itemize} 
          \item \citet{opaleos} + \citet{straniero1988} \citep[see also][]{prada2002}
     \end{itemize}
  \item Mass Loss:
      \begin{itemize} 
        \item Reimers 
      \end{itemize}
   \item External Boundary Conditions:
       \begin{itemize}
           \item  scaled solar $T(\tau)$ relation \citep{KS1966}
        \end{itemize}
   \item Heavy element distribution:
      \begin{itemize} 
        \item scaled solar \citep{lodders2009} 
      \end{itemize}      
        
\end{itemize}
 
 The Henyey method employed in the FuNS to solve the stellar structure equations is a first order implicit method and the accuracy of the solutions depends on the assumed mass and time resolutions. We adopt an adaptive algorithm to select the grid of mass shells and the time steps. In particular the variations between adjacent mesh points of luminosity, pressure, temperature, mass, and radius should be:
\begin{equation}
    0.8\times \delta < \frac{\left ( A_{N+1}-A_{N-1} \right )}{A_N}< \delta, 
\end{equation}
where A is one of $L, P, T, M$ or $R$.
In the present work we use $\delta=0.05$ everywhere, except for the shells located around some critical points, such as the boundaries of the convective zones, for which we use $\delta=0.005$. Similarly, the algorithm that we use to fix the time steps limits the variations of physical and chemical variables. Typically, when models climb the RGB, the total number of mesh points is about 1500, while the time step decreases from $\Delta t\sim 3\times 10^4$ yr, at $\log L/L_\odot=2.1$, down to $\Delta t \leq 10^3$ yr, for $\log L/L_\odot > 3$. 
 
 The set of stellar models on which our analysis is based is illustrated in table \ref{tab:modelli}.  These models share the age at the RGB tip, namely, $\sim 13$ Gyr, and the initial He mass fraction, namely, $Y=0.25$. The metallicity ($Z$) and the corresponding $[\rm{M}/\rm{H}]$ are listed in column 1 and 2, respectively, while the resulting bolometric magnitudes at the RGB tip are reported in column 4. Variations of the Cluster parameters, i.e., age, $Y$ and $Z$ as well as the theoretical uncertainties will be discussed in section \ref{sec:absolutem} and section \ref{sec:errteo}, respectively. 
 
 
 \subsection{Axions by RGB stars}\label{sec:axions}
 In stellar interiors, axions can be produced trough their interactions with standard model particles, mainly with photons and  electrons and, to a lesser extent, with nucleons. The Primakoff process, i.e., the conversion of a photon into an axion in the electromagnetic field of an ion, $\gamma + Ze \rightarrow a + Ze$, is the major consequence of the coupling with photons. 
 The energy-loss rate due to this process depends on the square of $g_{a\gamma}$, a quantity representing the strength of the axion-photon coupling. This rate increases as $T^4$, but it is suppressed at the density of the RGB core due to the electron screening that reduces the effective charge of the nuclei. By adopting the present upper bound for the axion-photon coupling, i.e., $g_{a\gamma}=6\times 10^{-11}$ GeV$^{-1}$ \citep{ayala2014,cast2017}, 
 we found that the Primakoff process alone would produce a maximum decrease of the tip bolometric magnitude by $\sim 0.05$ mag. 
 Regarding the coupling with electrons, the most important axion sources at $T\sim 10^8$ K are  the Compton scattering, $\gamma + e \rightarrow \gamma + e + a$, and the Bremsstrahlung, $e + Ze \rightarrow e + Ze + a$. As noted by \citet{raffelt1994},
 owing to the large electron degeneracy which develops in the core of a red giant star, the Compton process is suppressed due to the Pauli blocking. On the contrary the Bremsstrahlung process may induce a sizeable axion production. The energy-loss rate due to Bremsstrahlung axions scales as the square of $g_{ae}$, a quantity representing the strength of the axion-electron coupling. 
 
 In figure \ref{loss}, we compare the plasma-neutrino luminosity with those expected from axions emission due to Compton and Bremsstrahlung. This figure refers to a model with $M=0.82$ $M_\odot$, $Y=0.25$, $Z=0.001$, and an axion-electron coupling $g_{ae}=10^{-13}$. Almost the whole RGB evolution is shown, from $\log L/L_\odot=1$ to the tip. After the RGB bump,
 which corresponds to the double luminosity inversion occurring at $\log L/L_\odot \sim 2$, the neutrino emission becomes the dominant energy-loss process. Close to the RGB tip (just before the He ignition) the neutrino luminosity is about 5 times the axion luminosity. On the other hand, Bremsstrahlung is the leading axion emission process in RGB stars, more than 1 order of magnitude brighter than Compton. Finally, figure \ref{flash} shows the energy-loss rates within the core just before the He ignition. Note that while neutrinos are mainly emitted near the center, where the density is higher, the axion energy loss peaks at $\sim 0.2$ $M_\odot$ out of the center, where the temperature attains its maximum.    
 
\begin{figure}
   \centering
   \includegraphics[width=9.0cm]{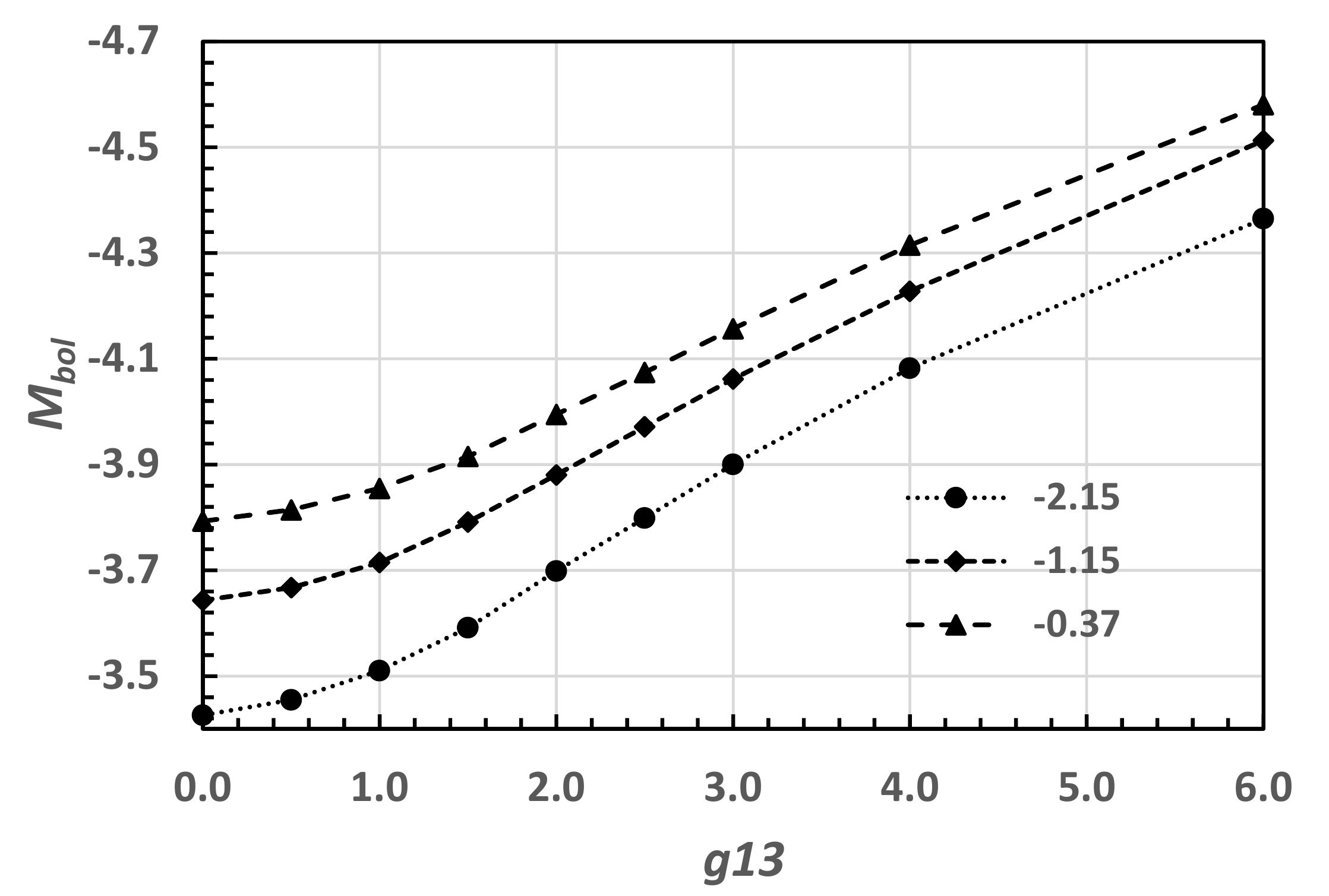}
   \caption{The relation $M_{bol}$ at the RGB tip versus $g_{13}$ for 3 different metallicities ($[\rm{M}/\rm{H}]$). }
   \label{rbol13}
\end{figure}

  The variations of the tip bolometric magnitudes as a function of $g_{ae}$ is reported  in table \ref{tab:modelli} for models with different initial metallicity (see also figure \ref{rbol13}). Here $g_{13}=g_{ae}\times 10^{13}$.
  The corresponding reductions of the tip bolometric magnitudes with respect to models without axions are reported in the last column of table \ref{tab:modelli}. Note that the more metal poor models show a higher sensitivity to the axion-electron coupling. As a whole, a $g_{13}=6$ would imply $\sim 1$ mag brighter bolometric magnitude, a value which is evidently incompatible with the observed RGB tip luminosity. A more stringent constraint can be obtained by means of a detailed evaluation of the error budget.    
 
  \begin{table*}
	\centering
	\caption{Uncertainties of the input physics adopted in the   calculations of  the tip bolometric magnitudes.}
    \label{tab:theounc}
	\begin{tabular}{lccc} 
		\hline 
		parameter &  reference & uncertainty & $\Delta M_{bol}$$^1$ \\
    \hline
  $\eta$ (mass loss) & Reimers &    $0.1 : 0.5$       & 0.014  \\
  ${14}$N+p  & STARLIB (LUNA) &       $\pm 10$\%      & 0.016  \\
  $3\alpha$  & STARLIB (NACRE) &       $\pm 20$\%     & 0.033  \\
  screning $3\alpha$ & Dewitt 1973 + Graboske 1973 + Itoh 1979 &   $\pm 20$\%          & 0.035  \\
  neutrinos      &  Haft 1994 + Itoh 1996 &       $\pm 5$\%     & 0.026  \\
  e conductuvity &  Potekin 1999 & $\pm 5$\%          & 0.049  \\
  rad. opacity   &  OPAL+COMA 2006   & $\pm 5$\%      & $<0.001$  \\
    $\alpha$ (mix. length)  & 1.82 (SSM calibrated)   & $1.62 : 2.02$  & $<0.001$\\
  boundary condition & $T(\tau)$ rel. (Krishna Swamy 1966)  & $\pm 10$\% & $<0.001$\\
  Eq. of state   &  OPAL 2005 + Straniero 1988 & see text & $0.003$  \\
  microscopic diffusion & Thoul et al. 1994  & see text & $0.006$ to $0.025$\\
 		\hline 
 \multicolumn{4}{l}{\footnotesize $^1$Full width variation of the tip bolometric magnitude.}\\		
       \end{tabular}
\end{table*}

\subsection{Theoretical uncertainty}\label{sec:errteo}
For the purpose of the present work, we have carried out a {\it sensitivity study}, with the aim to assign a sort of {\it theoretical error} to the model prediction of the RGB tip luminosity. There are many examples of this kind of study in the extant literature \citep{viaux2013_AA, valle2013, serenelli2017}. In these papers, the reader can find exhaustive illustrations of the various uncertainties affecting the relevant input physics. We do not repeat this discussion.
The uncertainties we have adopted are listed in table \ref{tab:theounc}. From column 1 to 4, we have reported: 1) the model inputs; 2) the reference for the preferred  values;  3) the assumed $1\sigma$ errors; 4) the variations of the RGB tip bolometric magnitudes implied by the variation of each input physics within the assumed error bar (full width). The quantities $\Delta M_{bol}$ reported in the last column represent the {\it sensitivity} of the predicted tip bolometric magnitude to the various model inputs. 
Note that the relationship between tip luminosity and model input is often non-linear, so that the uncertainty may be not perfectly symmetric around the central $M_{bol}$ value.  For this reason, in the last column of table \ref{tab:theounc} we have reported  the full-width variation rather than the half-width. In any case, a rough estimation of the cumulative theoretical error may be obtained by summing in quadrature the (average) half-width variations, i.e., $\Delta M_{bol}/2$. The result is  $\sigma_{theo}\sim0.04$.  
On the other hand, a more accurate estimation of this error can be obtained by means of a Monte-Carlo propagation method. In practice, we have assumed that each one of the listed model inputs (except for the microscopic diffusion and the equation of state) may vary according to a normal error distribution function. 
This assumption, which is appropriate when the uncertainty represents statistical fluctuations of experimental measurements, it is certainly not strictly appropriate when the model input has been obtained through a theoretical calculation. Nevertheless, this is still a reasonable and simple assumption when the theory can identify a preferred (or best) value for the model input and values far from it are less likely than values close to it. For example, the neutrino energy-loss rates are  computed on the base of the Weinberg-Salam electro-weak theory 
\citep[see, e.g.,][]{raffelt_book}. 
In the last 30 years, precision experiments have tested this quantum field theory at the level of 
one  percent  or  better. Nevertheless the neutrino rates depend on some constants, such as the Weinberg angle, not predicted by the theory, whose values are experimentally determined. This occurrence introduce a certain uncertainty that we assume to be modeled by means of a proper normal error distribution\footnote{The possible existence of a non-zero neutrino magnetic moment, that would  imply a systematic increases of the energy-loss rates, is not considered here (see section \ref{sec:conclusions}).}.

On the other hand, such an approach cannot be applied when the theory can not identify a preferred (or best) value of the model input, but a range of equally possible values. In that case a rectangular probability distribution is more appropriate. This is the case of microscopic diffusion and equation of state.

With equation of state (EoS), we intend the calculation of all the thermodynamical quantities that appear in the stellar structure equations. It includes the density, the specific heats, the adiabatic gradient and the partial derivatives of the density with respect to pressure, temperature and molecular weight. Note that in FuNS we use P and T as {\it native} thermodynamic variables. For $\log T<6.5$ we use the latest version of the OPAL EoS, while for larger temperatures we adopt the EoS described in \citet{straniero1988} \citep[see also][]{prada2002}. The latter assumes fully ionized matter and includes a detailed description of the relevant quantum-relativistic effects as well as the ion and electron Coulomb coupling. No approximate formulas are used for the evaluation of the Fermi-Dirac integral, which are always  computed numerically.  On the other hand, the OPAL EoS represents the state-of-the-art for the computation of the thermodynamical quantities in the partially ionized H-rich envelope. As a whole, the most important uncertainties affecting EoS calculations concern the deviations from a perfect gas that arise at high density. Then, to evaluate the sensitivity of RGB tip luminosity on the EoS, we have computed some stellar models by assuming a perfect gas EoS. Firstly, we have removed all the Coulomb corrections in the EoS calculation for $\log T> 6.5$. These corrections should mainly affect the core of  the brightest RGB models, where the central density approaches $\sim 10^6$ g/cm$^3$. In this case, the resulting tip bolometric magnitude is 0.0025 mag larger than that obtained with the reference models. Then, we have replaced the OPAL EoS with a perfect gas low-temperature EoS, in which the ionization degree has been obtained with a classical Saha equation. In this case, the resulting tip bolometric magnitude differs by less than $0.001$ mag with respect to the reference case. Hence, we have assumed a conservative systematic uncertainty associated to the EoS of $0.003$ mag. 

Concerning microscopic diffusion, the effect of the combined action of gravitational settling and thermal diffusion in main sequence stars is to slowly move  He and heavy elements toward the stellar center.  In the FuNS code microscopic diffusion is taken into account by inverting the set of Burgers equations \citep{thoul1994}. Note that in the evaluation of the Coulomb logarithm, full ionization is often assumed. Although this approximation is very common among the extant  calculations of evolutionary tracks and isochrones for globular cluster stars, it presents  some  limitations. For $\log T<6$, the Coulomb cross section is overestimated when full ionization is assumed and, in turn, the diffusion coefficients are underestimated.  In addition, the gravitational settling may be hampered by the  acceleration caused  by the  net  transfer of  momentum from the outgoing photon flux to ions and electrons, a process also known as radiative levitation \citep[][and references therein]{turcotte1998,delahaye2005}. 
Note that  the interior of a main sequence star with mass $\sim0.8$ $M_\odot$ is mostly stable against convection, with the exception of a small external convective zone, so that microscopic diffusion is the major instability affecting the internal composition. 
Owing to the large main sequence lifetime, this secular instability may produce sizeble modifications of the chemical stratification and, in turn, of the physical structure \citep{proffitt1991, chaboyer1992, straniero1997}. 
However, just after the central H exhaustion, the convective instability penetrates inward, so that about the 80\% of the stellar mass is homogenized (first dredge up). 
As a consequence, the original amounts of He and heavy elements are restored in a large part of the star and the evolutionary track resemble that of a model computed without diffusion. Later on, during the RGB phase, the evolutionary lifetime becomes much shorter than the diffusion timescale. Nonetheless, the star still keeps some memory of the modifications induced by diffusion in the previous evolutionary phase. Small changes in some critical quantities are found.
In particular, RGB models with diffusion present a larger core mass at the He flash and, in turn, a higher RGB tip luminosity.

In our reference models (table \ref{tab:modelli}) we have switched off the microscopic diffusion. Nonetheless, in our analysis we have considered its effect as a systematic uncertainty. In practice, to  maximize the effects of microscopic diffusion we have  neglected the radiative levitation. As a result, the tip bolometric magnitudes are from $-0.006$ mag (at Z=0.006) up to -0.025 mag (at Z=0.0001) brighter than that of the corresponding reference models (no diffusion). 

\begin{figure}
   \centering
   \includegraphics[width=9.0cm]{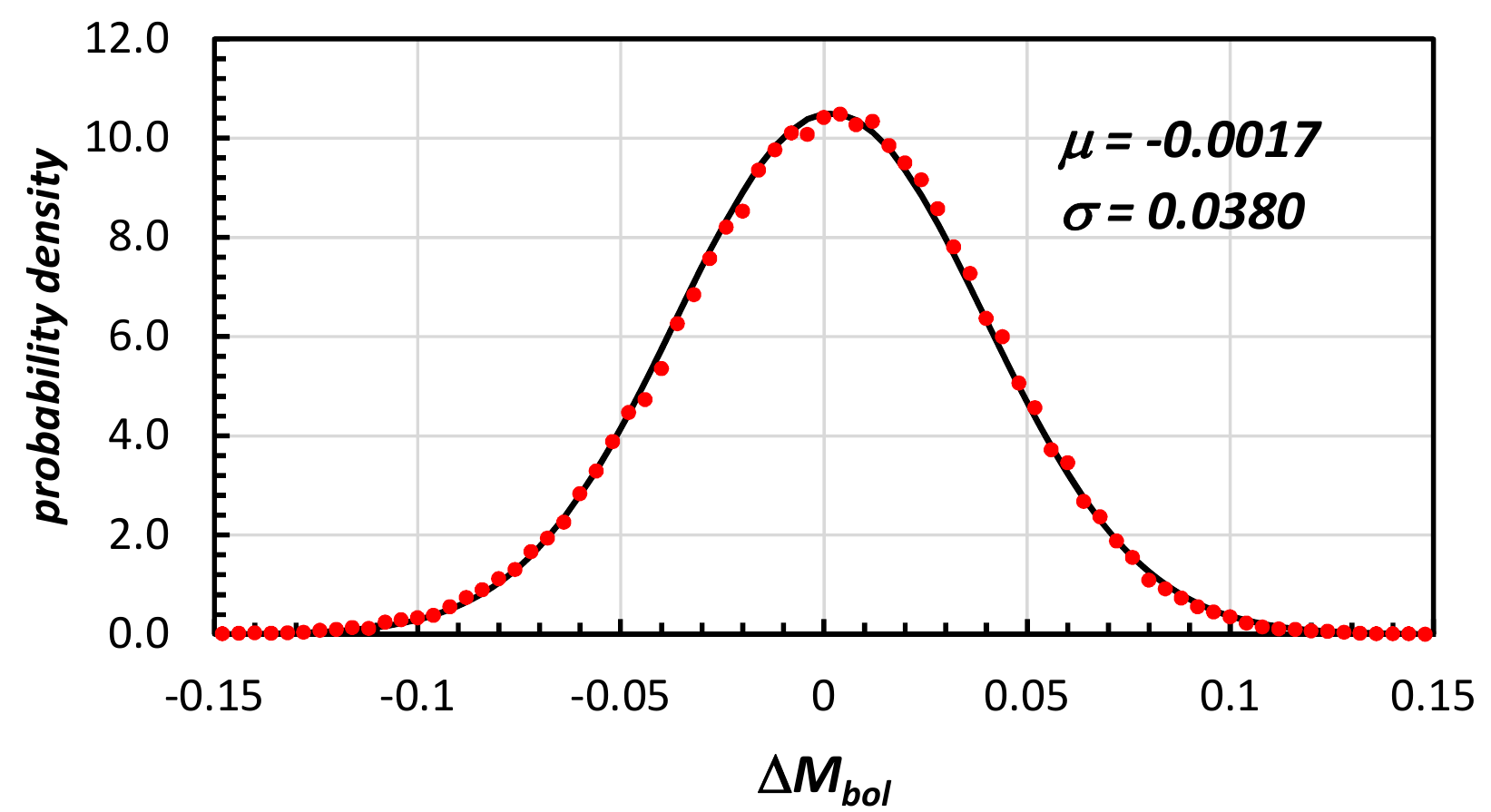}
   \caption{Result of the Monte-Carlo error propagation: points represent the normalized frequency of Monte-Carlo events, while the curve is the best fit Gauss function. Mean and standard deviation are shown. }
   
   \label{mc}
\end{figure}

Summarizing, in the Monte Carlo calculation, we have assumed that the uncertainties associated to EOS and microscopic diffusion produce equally probable variations in between the two extreme cases we have previously described.  For all the other model inputs we have assume a normal error distribution with $\sigma$ values as reported in column  3 of table \ref{tab:theounc}.
Some targeted checks have been done to verify that the variations of the tip luminosity implied by the variations of the various model inputs are uncorrelated. This occurrence allow us to evaluate the cumulative shift of the tip luminosity simply by summing the independent effects induced by the variations of the single model inputs. Then,  
we have calculated a large number ($10^5$) of tip $M_{bol}$ by randomly varying the model inputs. The result for an hypotetical Cluster with $age=13$ Gyr, $[M/H]=-1$ and Y=0.25 is illustrated in figure \ref{mc}. The normalized frequency histogram is well fitted by a Gaussian distribution with  mean $\Delta M_{bol}\sim 0$ and standard deviation $STD=0.038$.  Finally, we have repeated the Monte-Carlo calculation changing the Cluster parameters, i.e., the age and the initial composition, finding only marginal variations in the resulting standard deviation. 

Summarizing,  the adopted theoretical error of the tip bolometric magnitude  is $\sigma_{theo}=0.038$:
\

\begin{table*}
	\centering
	\caption{Comparisons with  previous calculations. The model parameters, i.e., mass, metallicity and He mass fraction are reported in the first three columns. The RGB tip luminosity  obtained with the FuNS is in column 4, while the corresponding differences with previous calculations (FuNS-previous calculations) are in column 5 to 8. The result obtained with the FuNS code, but imposing the weak nuclear screening everywhere and at any time (see text), is in the last column.}
    \label{tab:confronti}
	\begin{tabular}{ccccccccc}
	\hline
  M/M$_\odot$  &   Y   &   Z      &  $\log L/L_\odot$ (FuNS) & BASTI &  PGPUC  & GARSTEC &  PISA & weak sc (FuNS)\\
\hline
0.82 & 0.245 & 1.36E-03 &  3.337 &       &  -0.037 &         &   & -0.030  \\
0.80 & 0.246 & 1.00E-03 &  3.358 & 0.015 &     &  -0.022 &  -0.009 &  \\
0.80 & 0.245 & 1.00E-04 &  3.272 & 0.020 &     &  -0.022 &  -0.005 &  \\
\hline
\end{tabular}
\end{table*}

 \subsection{Comparisons with previous calculations}
 In the extant literature there exist several  papers reporting calculations of RGB stellar models for globular cluster stars. Here we have compared our results with the most recent ones, as obtained with 4 different evolutionary codes, namely: BaSTI \citep[][http://basti.oa-teramo.inaf.it]{basti2004}, PGPUC \citep{PGPUC2012,viaux2013_AA}, GARSTEC \citep{serenelli2017} and PISA \citep[][http://astro.df.unipi.it/stellar-models]{PISA2012}. To perform these comparisons we have computed specific models with the same initial stellar parameters as in the previous calculations. The results of this comparisons are reported in table \ref{tab:confronti}.
 In general, the differences in the tip bolometric magnitude are within the quoted $1\sigma$ theoretical uncertainty (see section \ref{sec:errteo}). However, a larger discrepancy is found with respect to the PGPUC model, the latter being about 0.09 mag fainter (or $\delta\log L/L_\odot=0.037$). Since this model is the same adopted by \citet{viaux2013} to constrain the axion-electron coupling, we have attempted to understand the origin of this discrepancy. This faintness of the PGPUC models at the RGB tip was already noted by \citet{serenelli2017}. These authors suggested that it is due to the evaluation of the screening enhancement factors which multiply the nuclear reaction rates. Indeed, \citet{viaux2013,viaux2013_AA} followed the prescriptions of \citet{salpeter1954}, who developed the theoretical framework for the calculation of the
 screening potential in the two extreme cases of {\it weak} and {\it strong} regime of the stellar plasma. In the weak screening regime, the coupling between a nucleus and the nearby electrons and nuclei is weak or, more precisely, the Coulomb interaction energy is much smaller than the thermal energy ($KT$). On the other hand, the strong regime refers to the case of strong coupling, i.e., Coulomb interaction energy much greater than the thermal energy. The weak regime condition is usually fulfilled in the core of H-burning stars with mass $\sim 0.8$ $M_\odot$. However, at the higher density of the core of a RGB star ($\sim 10^6$ g/cm$^3$), the interaction energy is comparable to the thermal energy and the weak screening prescription by \citet{salpeter1954} largely overestimates the screening factors.     
 As shown by \citet{dewitt1973,graboske1973}, the intermediate regime is more appropriate to describe the plasma conditions in the core of a RGB star. As an example, for a He-rich gas whose temperature and density are $10^8$ MK and $10^6$ g/cm$^3$ respectively, the effective $3\alpha$ rate is more then a factor of 2 larger when the weak screening is imposed. As a result, the He ignition should occur at lower luminosity. To quantify this effect, we have calculated an additional model, $M=0.82$ $M_\odot$, $Z=0.00136$ and $Y=0.25$, by imposing week screening factors for all the nuclear reactions, from the beginning of the evolution up to the RGB tip. As reported in the last column of table \ref{tab:confronti}, at the RGB tip this model is about 0.075 mag fainter than our reference model computed by calculating the screening factors according to  \citet{dewitt1973,graboske1973}. We conclude that most of the discrepancy between our models and the PGPUC ones is likely due to the different nuclear screening prescriptions. 
 As we will show in section \ref{sec:analysis}, the larger RGB tip luminosity we obtain by adopting the intermediate screening reduces the need of an additional energy loss from the RGB core and, in turn, it implies a lower upper bound for $g_{ae}$.   
 
 Let us finally mention a discrepancy between our sensitivity study and that discussed in \citet{valle2013}. At variance with our result, these authors find that the major uncertainty affecting the RGB tip luminosity is due to a variation of the radiative opacity. On the other hand, our result confirms previous finding reported by \citet{viaux2013} and \citet{serenelli2017}. In principle, the energy transport within the core of a RGB star is mainly driven by the heat conduction from degenerate electrons, so that a 5\% variation of the radiative opacity at high temperature is expected to produce a small effect on the tip luminosity. Outside the core, the radiative opacity affects the temperature gradient, which is super-adiabatic in a large portion of the convective envelope. As a result,  the effective temperature of a RGB star is modified by a change of the radiative opacity, but its luminosity is only marginally affected. For these reasons, although further investigations are certainly needed to clarify the origin of such a discrepancy, we are quite confident about the robustness of our prediction.


\section{Combining photometries from HST and ground based telescope} \label{sec:clusters}
 In this section we illustrate the strategy we have followed to derive the RGB tip bolometric magnitudes from optical (V and I) and near-infrared (J and K) photometries. 

\subsection{The VI  catalog}\label{sec:VI}
In order to select samples of RGB stars as complete as possible, the adopted strategy consisted in the combination of different catalogs characterized by complementary properties. In particular, to resolve the crowded central regions, we exploited the high spatial resolution of space based telescopes, while to sample the entire radial extension of the clusters, we used photometric data from ground based facilities that cover  larger field of view (FoV). To this aim, the required data sets  were extracted from public catalogs, namely: the  ``ACS globular cluster Survey'' \citep{sarajedini07} and the ``Homogeneous UBVRI photometry of globular clusters'' \citep{stetson19}. In particular, we used V and I magnitudes from the latter and   the magnitudes in the filters F606W and F814W, as calibrated into the ground photometric system, for the former. By using common stars, we have carefully checked that the ground system magnitudes tabulated into the HST catalog and those from the ground based catalogs were homogeneous. 

For the present work, we have selected three clusters, namely: M5, NGC 362 and 47 Tuc. The first is the cluster already used by \citet{viaux2013_AA} to constrain the strength of the axion-electron coupling, while the other two clusters are the only ones for which reliable parallax distances are available after the Gaia DR2 \citep{chen2018}. 
In the specific case of NGC 362, since the ground based catalog was not complete as for M5 and 47 Tuc, we have complemented our analysis with additional observations from the ESO archive, i.e., FORS2 data from the program 60.A-9203. In particular, we have analysed two short images (1 s each) obtained with the standard resolution (0''.25 /pixel for a total FoV of $6'.8\times6'.8$, hereafter SR) and two (1 s each) with high resolution (0''.125 /pixel for a total FoV of $4'.2\times4'.2$, hereafter HR) collimator  in both $I_{BESS}$ and $V_{BESS}$ filters.
For both datasets (i.e. HR and SR), the photometric analysis has been performed using DAOPHOT II \citep{stetsonDaophot}. Briefly, for each exposure we first modelled a spatially variable Point Spread Function by using more than 50 isolated and bright stars. We then applied this model to all the sources detected at more than $3\sigma$  from the background level  and we built the final catalog containing all the sources detected in both filters.
Finally, we have reported the instrumental magnitudes to the absolute reference system by cross-correlation\footnote{To perform the cross-correlation, we have used CataXcorr, a code aimed at cross-correlating catalogs and finding astrometric solutions, developed by P. Montegriffo at INAF
- OAS Bologna. This package has been successfully used in a large number of papers in the past years (e.g. see \citet{pallanca13,pallanca14,pallanca19,cadelano17,Dalessandro14}).} with the Standard Stetson catalog.
Then, the combined catalogs were built as the combination of the different available catalogs.
When available, we have preferentially adopted the HST magnitude, not prone to crowding. However, for a few central very bright stars, we have preferred to anyway adopt the ground based magnitudes, since the HST magnitudes may be  biased because of the non linear regime close to the photometric saturation level.  Note that for such bright objects the crowding is not a problem even in the most central regions.
In the case of NGC 362, for which also two FORS2 data sets were available, the priority adopted in the catalog compilation has been HST, FORS2 HR, FORS2 SR and Stetson standard but for very bright objects, for which we have adopted the FORS HR magnitudes even if detected in the HST catalog. Finally, we have dereddened the magnitudes by adopting the $E(B-V)$ values tabulated in \citet[][2010 version]{harris96} and adopting a standard extinction law with $Rv=3.1$. In particular, we have assumed ${\rm A_V}=3.12\ {\rm E(B-V)}$ and ${\rm A_I}=1.87\ {\rm E(B-V)}$ \citep{cardelli89, odonnell94}.

The apparent bolometric magnitudes of the RGB stars have been calculated by means of the UBVRIJHK color-temperature calibration reported in \citet{worthey11}. In all cases, we have assumed $\log {\rm g} =0$, which is a typical gravity value for bright RGB stars. Then, we have estimated the bolometric correction of all stars making use of their own dereddened color ${\rm(V-I)_0}$.
The combined errors on the bolometric magnitudes have been calculated as the propagation of several sources of uncertainty, namely:
the photometric error as estimated with DAOPHOT, all the sources of errors in the determination of the calibrated and dereddened magnitudes (e.g. the value of reddening $\rm {E(B-V)}$, the extinction law, the calibration zero points, etc...) and the error on the bolometric correction. The latter includes the intrinsic uncertainty of the BC-color empirical relation by \citet{worthey11} and that of the measured V-I colors. For the brightest RGB stars, this uncertainty is of the order of $\pm 0.15$ mag and it represents the major contribution to the $m_{BOL}$ error budget. 

\subsection{The near IR catalog}\label{sec:IR}
In addition to the VI photometry described in the previous section, we have also considered the 
near-IR catalog of GGCs presented in a series of paper by the Bologna team \citep{ferraro2000,valenti2004a}.
Basically, the catalog contains a set of JHK photometric data of 22 GGCs, obtained at the  
ESO MPI 2.2 m telescope equipped with the near-IR camera IRAC-2. These data cover the most central $4\times 4$
arcmin$^2$ regions of each cluster. In some cases, additional data from the 2mass survey have been used to  sample
more external regions \citep{valenti2004a}. For five clusters, i.e., M3, M5, M10, M13 and M92,
 we have also considered the J and K observations obtained by \citet{valenti2004c} with the
Telescopio Nazionale Galileo (TNG) equipped with the ARNICA IR camera. These observations  
cover the most crowded $1.5\times 1.5$ arcmin$^2$ central regions. Finally, a quite large sample of giants belonging to the RGBa of  $\omega$ Cen, as obtained by \citet{sollima2004} with the ESO NTT telescope during the commissioning of the SOFI camera,  has been also used. These dataset covers  a $13\times 13$ arcmin$^2$ area around the cluster center. Then, the bolometric magnitudes have been estimated by means of the empirical bolometric corrections derived by \citet{buzzoni2010}. In particular, we have
used their BC$_K$ versus (J-K) color relation. 
As for the VI photometriic samples, the errors of the bolometric magnitudes were calculated by combining the uncertainties due to the photometry with those affecting the bolometric correction. 
In general, the K-band bolometric corrections are more reliable than the V and I ones. According to \citet{buzzoni2010}, for the brightest RGB stars, we have assumed $\sigma_{BC}=\pm 0.1$ mag.  
 
 \subsection{Selection of RGB  stars}\label{sec:selrgb}
Particular care has been paid to decontaminate the RGB photometric samples from background field stars and from the presence of AGB stars. The selection of RGB star candidates has been done basing on the available colors and magnitudes (VIJHK). Given the very small photometric errors  of RGB and AGB stars ($\sim 0.01$ mags) and the expected color separation between the two parent sequences
($0.02<$(V-I)$_{RGB}-$(V-I)$_{AGB}<0.1$), we found a clear separation between the two populations from a minimum of 2$\sigma$, around 1 mag below the RGB-tip, up to more than 20$\sigma$ at larger magnitudes. A good example of this RGB-AGB separation  can be found in Figure 1 of \citet{beccari2006}, where the two populations of 47 Tuc, thanks to the
high quality of the data, which is similar to that of the present work, are well populated and clearly distinguishable.
Such a clear separation, allowed us to select the RGB sample through a box selection drawn around the RGB fiducial line and properly checked by a visual inspection.

 
\subsection{The RGB tip luminosity from GGC photometries} \label{sec:tipobs}
\begin{figure}
   \centering
   \includegraphics[width=9.0cm]{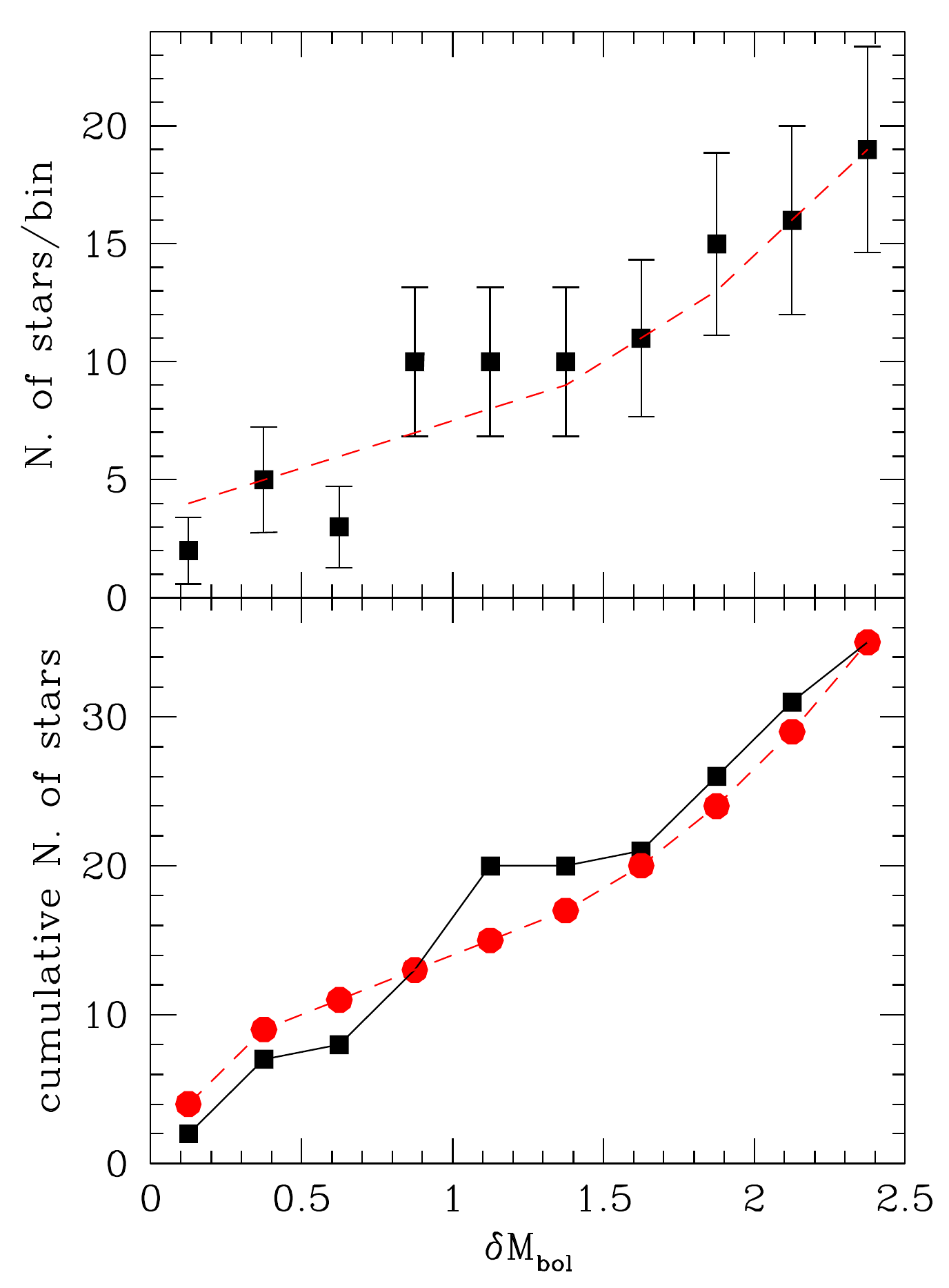}
   \caption{Upper panel: the observed luminosity function of the 2.5 mag brightest portion of the RGB of the cluster  M5 (squares) is compared to the theoretical luminosity function (dashed line). Poisson errors ($\sqrt{N}$) are also shown. Lower panel:
   the cumulative LF of the brightest portion of the M5 RGB (solid line and squares) is compared to the theoretical expectation (dashed line and circles).}
   \label{lf}
\end{figure}

\begin{figure}
   \centering
   \includegraphics[width=9.0cm]{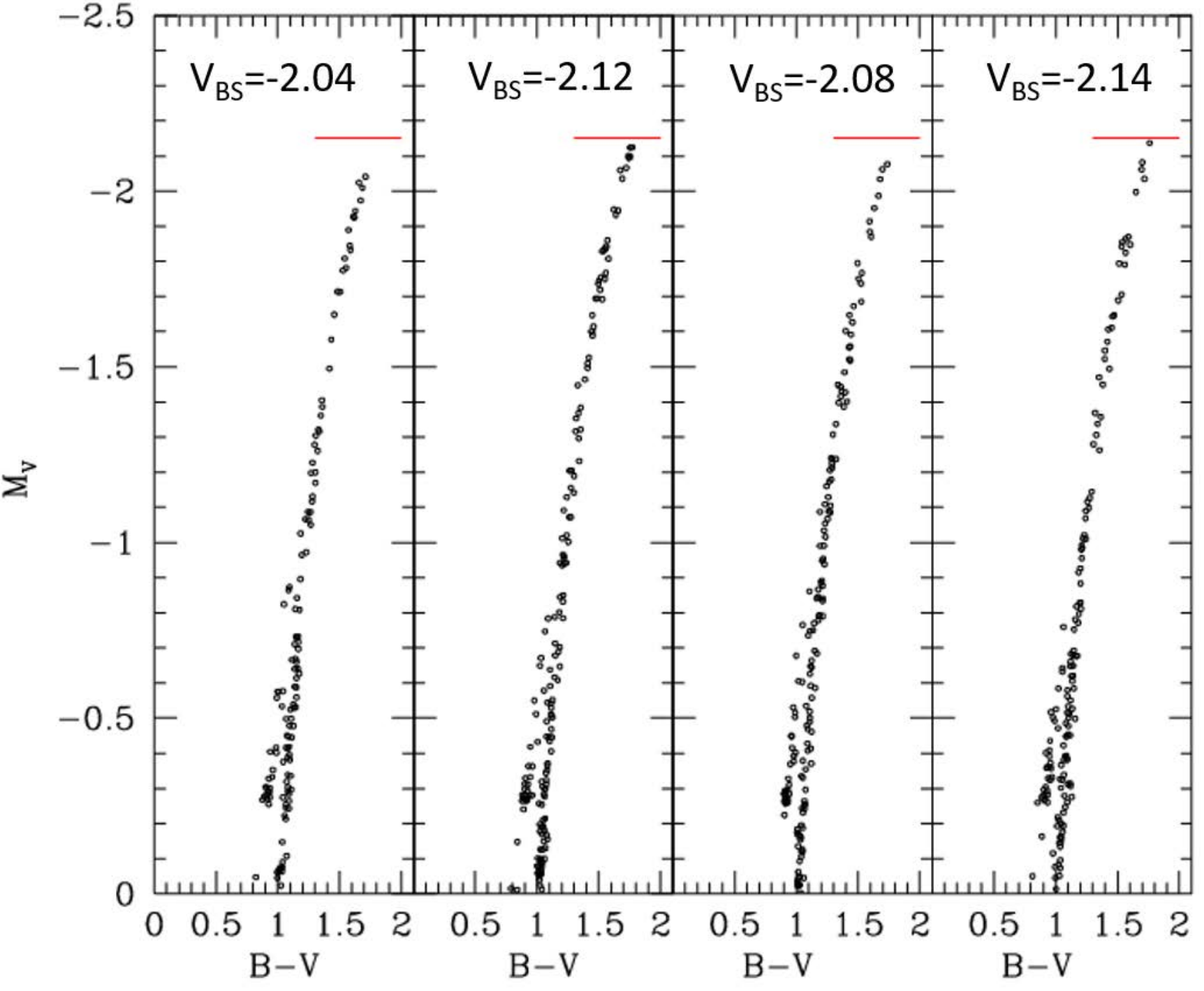}
   \caption{Synthetic color-magnitude diagrams of bright globular cluster stars. All the SCMDs have been computed with the same input parameters, in particular, same age, chemical composition  and number of stars. The location of the RGB tip is marked by a red horizontal line. For all the SCMDs it is at $M_V=-2.15$. The $M_V$ mag of the brightest stars are reported in each panel (V$_{BS}$). }
   \label{scmd}
\end{figure}

     In this section we illustrate the method we follow to determine the luminosity of the RGB tip from a finite photometric sample of RGB stars. This problem is equivalent to the evaluation of the difference of the luminosity of the RGB tip and that of the brightest observed RGB star. To do that, we will make use of theoretical luminosity functions (LFs) and synthetic color-magnitude diagrams (SCMDs). First of all, let us verify that the observed and the theoretical LFs are drawn from the same distribution. In figure \ref{lf}, the observed luminosity function of the cluster M5 is compared to the theoretical expectation. By means of a classical Kolmogorov-Smirnov test we found a statistic parameter d=0.2 and a probability of 97.5\% that the theoretical and the observed data sets are drawn from the same distribution. 
    Furthermore, a chi-squared test leads to  a 86\% probability that the theoretical and the observed LFs are  drawn from the same distribution.
     These statistical tests support our use of the theoretical LFs to evaluate the differences between observed luminosity of the brightest RGB stars and the RGB tip luminosity. 
    Then, synthetic color-magnitude diagrams are generated following a standard Monte-Carlo procedure. Some examples are illustrated in figure \ref{scmd}. In computing these four SCMDs we have assumed the same input parameters, in particular, same age, chemical composition, B and V photometric uncertainties and total number of stars. In all cases, the RGB tip occurs at M$_V=-2.15$ mag. In contrast, the location of the brightest star changes randomly because of statistical fluctuations. 

 \begin{figure}
   \centering
   \includegraphics[width=9.0cm]{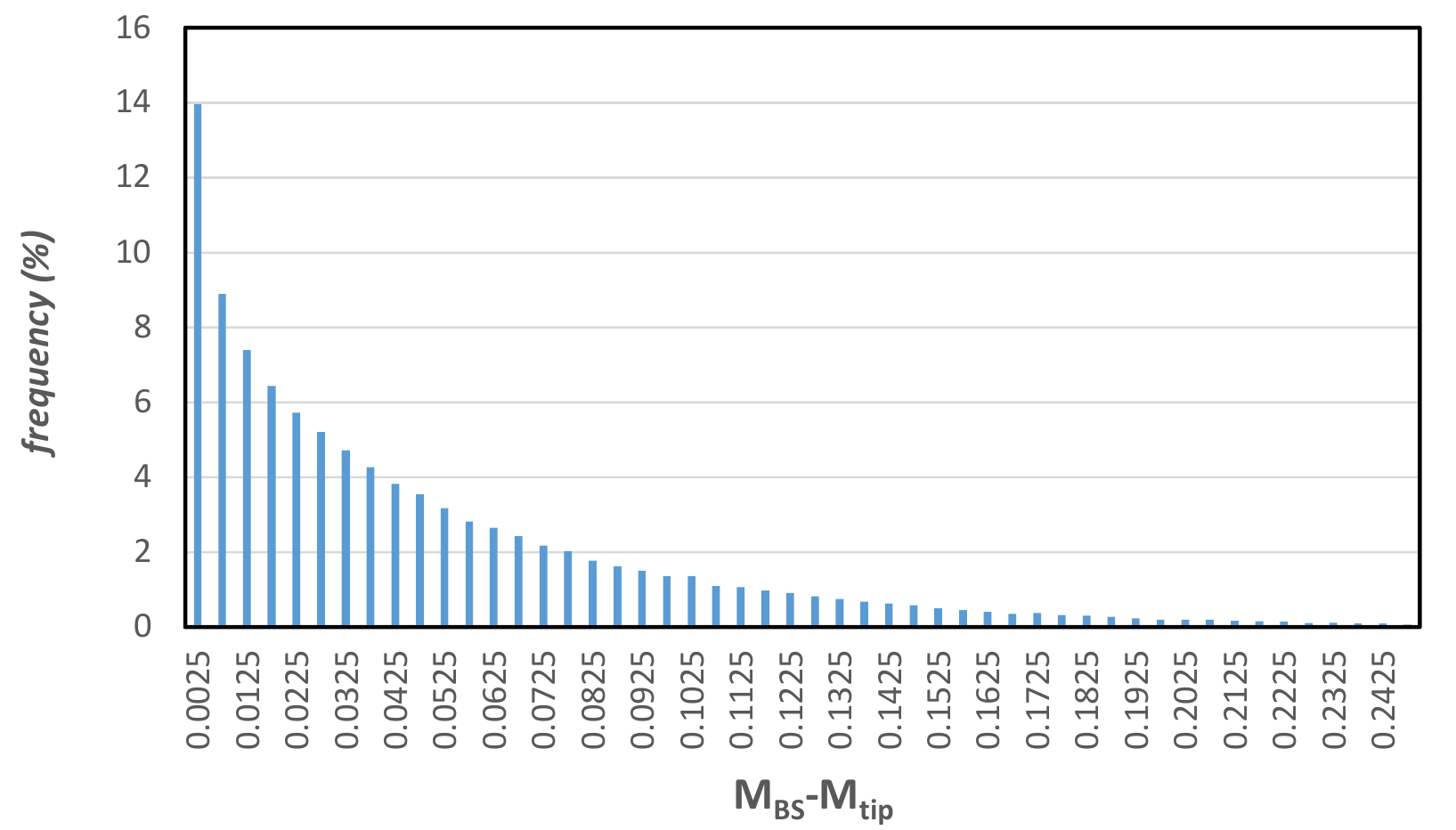}
   \caption{Histogram representing the frequency of the differences between the bolometric magnitute of the brightest star ($M_{BS}$) and that of the RGB tip ($M_{tip}$) for a sample of $10^5$ SCMDs computed for an age of 13 Gyr, Z=0.001, Y=0.25 and with 100 stars within  2.5 mag from the RGB tip}. 
   \label{diff}
\end{figure}

The analysis of these fluctuations can be done by collecting a sufficiently large number of these SCMDs, as illustrated in figure \ref{diff}. This frequency histogram represents the distribution of the differences between the bolometric magnitude of the brightest stars and that of the RGB tip in a sample of $10^5$ SCMDs. Each SCMD has been computed assuming an age of 13 Gyr, Z=0.001 and Y=0.25. In addition, each SCMD contains 100 stars in the brightest 2.5 mag portion of the RGB. The mode of the probability distribution is evidently 0, while the median is $<\delta>_{100}=0.030$ mag and the standard deviation from the median is $STD=0.056$ mag. 
However, for probability distributions with a long right tail,   the standard deviation overestimates the dispersion. For this reason, the mean deviation from the median (MDM\footnote{$MDM_N=\sum_{i}\frac{\left | \delta_i-<\delta>_N \right |}{n}$, where $\delta_i$ is the brightest-star to tip difference for the $i$ SCMD, $n$ is the total number of SCMDs and $N$ is the number of stars in the brightest 2.5 mag of the RGB.}),  or the interquartile range, are often preferred to the STD. For the probability distribution shown in figure \ref{diff}, we find $MDM_{100}=0.035$ mag, while the first and the third quartiles  are at $-0.019$ and $+0.033$ with respect to the median. Note that the semi-interquartile range is 0.026 mag, which is slightly smaller than MDM and about half the STD.
In table \ref{tip_corr}, we report the values of  medians, standard deviations, and mean deviations from medians, as obtained under different assumptions about age, chemical composition and the parameter $N$ that represents the number of stars in the brightest portion of the RGB. It results that the statistical descriptors  are practically insensitive to variations of age, Y and Z, within the range expected for globular cluster stars. In contrast, they are substantially affected by the number of bright RGB stars. 
 Then, knowing, for each cluster,  the bolometric magnitude of the brightest RGB star ($m_{BS}$) and the number of stars in the brightest 2.5 mag of the observed  RGB ($N$), we can estimate the tip bolometric magnitude as: $m_{tip}=m_{BS}-<\delta>_N$, where $<\delta>_N$ is the median of the distribution function of $m_{BS}-m_{tip}$.  
Concerning the total error, we have assumed $\sigma_{tip}^2=\sigma_{BS}^2+MDM^2$, where $\sigma_{BS}$ is the error on the apparent bolometric magnitude of the brightest RGB star (see sections \ref{sec:VI} and \ref{sec:IR}) and $MDM_N$ is the mean deviation from the median. Note that, in principle,  $m_{BS}-m_{tip}$ can not be negative, so that the error bar may be asymmetric. However,  the assumption of a symmetric error, although  it implies a small overestimation of the left side of the uncertainty,  has a negligible impact on the final results.
For practical purpose, good approximations of  median, standard deviation and  mean deviation from the median can be  obtained by means of the following equations (valid for $50\le N\le 300$):

\begin{equation*}
<\delta>_N = 0.03824 \times x^3 - 0.1627\times x^2 +0.09774\times x  + 0.179
\end{equation*}
\noindent
\begin{equation*}
STD_N = 0.07502\times x^3 + 0.57441\times x^2 - 1.51546\times x  +1.390
\end{equation*}
\noindent
\begin{equation*}
MDM_N= -0.07664\times x^3 + 0.56869\times x^2 - 1.43798\times x + 1.24946
\end{equation*}

\noindent
where $x=\log (N)$. Note that the photometric samples of all the 22 GGCs we have analyzed contain between 50 and 200 stars in the brightest 2.5 mag of the RGB. Clusters with $N<50$ have been excluded.  

\begin{table}
	\centering
      \caption{Medians, standard deviations and mean deviations from the median of the corrections applied to the bolometric magnitude of the brightest RGB stars. }
         \label{tip_corr}
		\begin{tabular}{cccccccl} 
		\hline
            \noalign{\smallskip}
            Age (Gyr) &  Z  & Y & N$^1$  & median & STD & MDM\\
            \noalign{\smallskip}
            \hline
            \noalign{\smallskip}
            13  & 0.001 & 0.25 & 50 & 0.063 & 0.107  & 0.072   \\
             13  & 0.001 & 0.25 & 100 & 0.030 & 0.056  & 0.035  \\
              13  & 0.001 & 0.25 & 200 & 0.008 & 0.030 & 0.018   \\
             13  & 0.001 & 0.24 & 100 & 0.030 & 0.056  &  0.035 \\
             13  & 0.0001 & 0.25 & 100 & 0.028 & 0.063 &  0.037  \\
             11   & 0.001 & 0.25 & 100 & 0.030 & 0.055 &  0.035    \\
       \hline
        \multicolumn{6}{l}{\footnotesize $^1$Number of stars in the brightest 2.5 mag of the RGB.} 
  \end{tabular}
\end{table}


\begin{table}
	\centering
	\caption{Uncertainties  of the cluster parameters and their effects on the tip bolometric magnitude.}
    \label{tab:clusterparam}
	\begin{tabular}{lccc} 
		\hline 
		parameter & reference value & uncertainty & $\delta M_{bol}$$^1$ \\

          \hline 
   Y              &  0.25   &   $ -0.01; +0.05$      & 0.045 \\
   $[\rm{M}/\rm{H}]$        & C2009$^2$   &   $   \pm 0.2$       & 0.080 \\
   Age            &  13 Gyr   &   $   \pm 10$\%   & 0.006  \\
          \hline
    \multicolumn{4}{l}{\footnotesize $^1$Full width variation of the tip bolometric magnitude.}\\      
    \multicolumn{4}{l}{\footnotesize $^2$\citet{carretta2009}. See text for more details.}\\      
  \end{tabular}
\end{table}

\subsection{Cluster parameters: distance, age and chemical composition.}\label{sec:absolutem}

Comparisons of models with observations require the  knowledge  of stellar parameters, such as the mass and the initial chemical composition.  As it is well known, globular cluster stars show substantial deviations from a scaled-solar composition. In particular, $\alpha$ elements, such as O, Mg, Si and Ca, are enhanced ($[\alpha/Fe]>0$). According to \citet{salaris1993}, reliable predictions of the observed stellar properties of $\alpha-$enhanced globular cluster stars, such as the RGB tip luminosity, can be obtained with models computed by assuming a scaled-solar composition with global metallicity\footnote{As usual, the global metallicity ($[\rm{M}/\rm{H}]$) and the total mass fraction of the elements with atomic number $\geq 6$ ($Z$) are related by $[M/H]=\log(Z/X)-\log(Z/X)_\odot$}:
\begin{equation}\label{eq:met}
    \left [ \frac{\rm{M}}{\rm{H}} \right ]=\left [ \frac{\rm{Fe}}{\rm{H}} \right ]+\log(a\times f_\alpha + b), 
\end{equation}
\noindent
where $f_\alpha=10^{\left [  \alpha/Fe \right ]}$ is the $\alpha$-enhancement factor, while $a$ and $b$ are numerical coefficients that depends on the adopted solar composition. According to the extant spectroscopic analysis of globular cluster stars, the average $\alpha$-enhancement is $[\alpha/\rm{Fe}]\sim 0.3$, with the most metal-rich clusters showing lower values and the most metal-poor presenting larger enhancements \citep{carney1996,pritzl2005}. In the present work, we have adopted the $[\rm{Fe}/\rm{H}]$ from \citet{carretta2009} and  we have assumed $[\alpha/\rm{Fe}]=0.4$, for clusters with $[\rm{Fe}/\rm{H}]<-2$,  $[\alpha/\rm{Fe}]=0.3$, for those with $-2<[\rm{Fe}/\rm{H}]<-0.8$, and  $[\alpha/\rm{Fe}]=-0.375[\rm{Fe}/\rm{H}]$, for the more metal rich. Since we have adopted the \citet{lodders2009} solar composition, the numerical coefficients should be: $a=0.6695$ and $b=0.3305$  \citep[see][]{piersanti2007}. Owing to the large uncertainty of the metallicity estimation, we have assumed a conservative $[\rm{M}/\rm{H}]$ error bar of $\pm0.2$. As reported in table \ref{tab:clusterparam}, the propagation of this error into the tip bolometric magnitude produces an uncertainty of about $\pm 0.04$ mag. The estimated $[\rm{M}/\rm{H}]$ value of the 22 clusters in our sample are listed in column 2 of table \ref{tab:clusters}.

Concerning the cluster age we have assumed $13\pm 2$ Gyr. Then, for each metallicity, we have selected the initial stellar mass corresponding to the assumed age. Note that this uncertainty of the age/mass implies a marginal variation of the estimated tip bolometric magnitude (see table \ref{tab:clusterparam}).  

For the bulk of the halo stellar population we have assumed an He abundance $Y=0.25\pm0.01$ \citep{izotov2014, aver2015J}. On the other hand, the possible presence of He-enhanced stars in the cluster samples should be also considered. According to
\citet{milone2018}, the He enhancement of second generation stars in the clusters of our sample is always smaller than $\delta Y_{2G-1G}<0.03$, while the maximum Y variation is generally smaller than $\delta Y_{max}<0.05$\footnote{A larger maximum variation (0.081) is quoted for NGC 6461, but it is affected by a large uncertainty ($\pm0.22$).}. Note that the RGB tip of He-enhanced stars is fainter than that of stars with normal initial He abundance. In addition, if the age difference between the first and the second generation of stars is negligible, the second generation RGB stars, due to the higher initial-He content, should have smaller masses. This occurrence partially counterbalances the decrease of the tip luminosity caused by the He enhancement in second generation stars. To quantify this effect, we have calculated a few He-enhanced models by reducing the initial mas in order to keep their age equal to that of normal-He stars (13 Gyr). In practice, assuming a maximum He variation among the stars of the same cluster of $\delta Y=0.05$, it results that the RGB tip of the most He-enhanced stars would be 0.037 mag fainter than that with normal He. Therefore, if the stars of the first generation are not a minority,  the brightest RGB stars are those with normal He. However, in case of an overwhelming majority of He-enhanced stars, higher statistical fluctuations of the observed tip luminosity are possible. In order to take into account such a possibility, in the analysis discussed in the next section, we will include an additional error affecting the tip magnitude due to the Y variations here illustrated. As reported in table \ref{tab:clusterparam}, our conservative estimation of this error is $\sigma_Y =0.045$, which corresponds to the systematic shift of the tip luminosity when the He content changes from $Y=0.24$ to $Y=0.30$ (or $\delta Y=0.06$). 

Comparisons of theoretical and observed luminosities also require the knowledge of the cluster distances. Among the classical methods used to derive the distance to clusters, widely exploited are those based on the location in the color-magnitude diagram of the  horizontal branch \citep[see, e.g.,][]{harris1996, ferraro1999, harris2010}. In particular, the mean luminosity of the horizontal branch or that of the zero age horizontal branch (\rm{ZAHB}) are often adopted as calibrated candles.  Since the horizontal branch luminosity depends on the chemical composition, either empirical or theoretical  $M_V(\rm{ZAHB})-\rm{metallicity}$ relations are used. Several alternative methods have been also investigated, such as those based on main-sequence or white-dwarf-sequence fitting, pulsation properties of RR Lyrae stars or orbital parameters of eclipsing binaries, even if their application is often limited to smaller sub-samples of the GGCs \citep[see][for a complete review of these methods]{harris2010}).
Recently, there has been a great expectation on the possibility of obtaining very precise parallax measurements with the Gaia astrometric satellite \citep{pancino2017}. After the second data release (Gaia DR2), however, it has been realized that for faint  sources (G > 14 mag) the Gaia parallaxes are affected by significant systematic errors. In particular, these systematics substantially affect the derivation of distance to GGCs. Nevertheless,
\citet{chen2018} (CH2018) were able to obtain quite accurate distances of two GGCs, NGC 362 and 47 Tuc, taking advantage of the background stars in the Small Magellanic Cloud and quasars to account for parallax systematics. Moreover, an indirect distance estimations of a large sample of GGCs has been recently obtained by \citet{baumgardt2019} (B2019). Combining proper motion and radial velocity measurements from the GAIA DR2 with ground-based measurements of line-of-sight velocities,  they obtained distances by fitting N-body GGC models to the observed kinematic properties. 

 In our analysis, we will make use of the distance moduli we have obtained  following the method described in \citet{ferraro1999} to determine the observed ZAHB luminosity. However, at variance with \citet{ferraro1999}, we have used a semi-empirical $M_V(\rm{ZAHB})-\rm{metallicity}$ relation. In practice, we have adopted  the relation reported in equation 4 of \citet{ferraro1999} to describe the quadratic dependence of $M_V(\rm{ZAHB})$ from the metallicity, but the zero-point has been re-calibrated in such a way to reproduce the absolute $M_V(\rm{ZAHB})$ of NGC 362 and 47 Tuc, for which precise distances from GAIA parallaxes are available. A weighted average of the 2 zero-points has been adopted, namely, $1.03 \pm 0.07$ mag. 

The distance moduli of the 22 clusters in our sample are reported in  table \ref{tab:clusters}. In particular, in column 3 there are the distance moduli based on the observed ZAHB luminosity, while in column 4 and 5 there are those based on Gaia parallax and kinematic measurements, as obtained by \citet{chen2018} and \citet{baumgardt2019}, respectively. Note that for the ZAHB distances, we have assumed a conservative common error of $\pm0.2$ mag, even if it may be smaller for some cluster. It includes the uncertainties in the determination of the ZAHB V magnitudes and that due to the adopted zero point of the $M_V(\rm{ZAHB})-\rm{metallicity}$ relation.


\begin{table*}
	\centering
      \caption{RGB tip bolometric magnitudes as obtained under different assumtion for the distance scale.}
         \label{tab:clusters}
		\begin{tabular}{lccccccc} 
\hline
         &      $[\rm{M}/\rm{H}]$     &    m-M (ZAHB)   &   m-M (CH2018)       &   m-M (B2019)      &   $M_{bol}$ (ZAHB)     &   $M_{bol}$ (CH2018)     &   $M_{bol}$ (B2019)    \\
\hline
 \multicolumn{8}{c}{near-IR photometry}\\          
\hline
M92      & $-2.05\pm 0.20$ & $14.76\pm 0.20$ &                 & $14.64\pm 0.08$ & $-3.63\pm 0.25$ & $             $ & $-3.51\pm 0.16$ \\
M15      & $-2.03\pm 0.20$ & $15.14\pm 0.20$ &                 & $15.05\pm 0.03$ & $-3.53\pm 0.24$ & $             $ & $-3.44\pm 0.13$ \\
M68      & $-1.97\pm 0.20$ & $15.14\pm 0.20$ &                 & $             $ & $-3.34\pm 0.32$ & $             $ & $             $ \\
M30      & $-2.03\pm 0.20$ & $14.73\pm 0.20$ &                 & $14.48\pm 0.24$ & $-3.65\pm 0.28$ & $             $ & $-3.41\pm 0.31$ \\
M55      & $-1.71\pm 0.20$ & $13.84\pm 0.20$ &                 & $13.63\pm 0.09$ & $-3.66\pm 0.25$ & $             $ & $-3.44\pm 0.18$ \\
$\omega$ Cen    & $-1.42\pm 0.20$ & $13.67\pm 0.20$ &          & $13.60\pm 0.02$ & $-3.59\pm 0.23$ & $             $ & $-3.52\pm 0.12$ \\
NGC 6752 & $-1.33\pm 0.20$ & $13.17\pm 0.20$ &                 & $13.15\pm 0.05$ & $-3.63\pm 0.25$ & $             $ & $-3.61\pm 0.16$ \\
M13      & $-1.36\pm 0.20$ & $14.43\pm 0.20$ &                 & $14.15\pm 0.10$ & $-3.56\pm 0.27$ & $             $ & $-3.28\pm 0.20$ \\
M3       & $-1.28\pm 0.20$ & $15.02\pm 0.20$ &                 & $14.88\pm 0.10$ & $-3.59\pm 0.24$ & $             $ & $-3.45\pm 0.17$ \\
NGC 362  & $-1.08\pm 0.20$ & $14.64\pm 0.20$ & $14.66\pm 0.12$ & $14.82\pm 0.07$ & $-3.48\pm 0.24$ & $-3.50\pm 0.17$ & $-3.66\pm 0.14$ \\
M5       & $-1.11\pm 0.20$ & $14.38\pm 0.20$ &                 & $14.40\pm 0.04$ & $-3.60\pm 0.25$ & $             $ & $-3.61\pm 0.16$ \\
NGC 288  & $-1.10\pm 0.20$ & $14.73\pm 0.20$ &                 & $15.03\pm 0.07$ & $-3.77\pm 0.24$ & $             $ & $-4.07\pm 0.16$ \\
M107     & $-0.81\pm 0.20$ & $13.94\pm 0.20$ &                 & $13.86\pm 0.14$ & $-3.55\pm 0.32$ & $             $ & $-3.47\pm 0.28$ \\
NGC 6380 & $-0.29\pm 0.20$ & $14.65\pm 0.20$ &                 & $             $ & $-3.97\pm 0.24$ & $             $ & $             $ \\
NGC 6342 & $-0.36\pm 0.20$ & $14.54\pm 0.20$ &                 & $             $ & $-3.76\pm 0.27$ & $             $ & $             $ \\
47 Tuc   & $-0.55\pm 0.20$ & $13.25\pm 0.20$ & $13.24\pm 0.06$ & $13.24\pm 0.02$ & $-3.78\pm 0.23$ & $-3.77\pm 0.13$ & $-3.77\pm 0.12$ \\
M69      & $-0.43\pm 0.20$ & $14.59\pm 0.20$ &                 & $             $ & $-3.53\pm 0.24$ & $             $ & $             $ \\
NGC 6441 & $-0.32\pm 0.20$ & $15.56\pm 0.20$ &                 & $15.36\pm 0.03$ & $-3.96\pm 0.24$ & $             $ & $-3.77\pm 0.13$ \\
NGC 6624 & $-0.31\pm 0.20$ & $14.55\pm 0.20$ &                 & $14.21\pm 0.13$ & $-3.90\pm 0.26$ & $             $ & $-3.56\pm 0.21$ \\
NGC 6440 & $-0.15\pm 0.20$ & $14.37\pm 0.20$ &                 & $             $ & $-4.00\pm 0.24$ & $             $ & $             $ \\
NGC 6553 & $-0.12\pm 0.20$ & $13.36\pm 0.20$ &                 & $14.19\pm 0.08$ & $-3.93\pm 0.25$ & $             $ & $-4.76\pm 0.17$ \\
NGC 6528 & $0.07 \pm 0.20$ & $14.24\pm 0.20$ &                 & $             $ & $-4.07\pm 0.24$ & $             $  \\                 
\hline                                                                          
\multicolumn{8}{c}{VI photometry}\\ 
\hline
NGC 362  & $-1.08\pm 0.20$ & $14.64\pm 0.20$ &  $14.66\pm 0.12$  & $14.82\pm 0.07$ & $-3.51\pm 0.26$ & $-3.53\pm 0.21$ & $-3.69\pm 0.17$ \\
M5       & $-1.11\pm 0.20$ & $14.38\pm 0.20$ &                   & $14.40\pm 0.04$ & $-3.63\pm 0.26$ & $             $ & $-3.64\pm 0.18$ \\
47 Tuc   & $-0.55\pm 0.20$ & $13.25\pm 0.20$ &  $13.24\pm 0.06$  & $13.24\pm 0.02$ & $-3.80\pm 0.26$ & $-3.79\pm 0.17$ & $-3.79\pm 0.16$ \\

\hline 
 \end{tabular}
 \end{table*}

\section{Results}\label{sec:analysis}
\begin{figure}
   \centering
   \includegraphics[width=9.0cm]{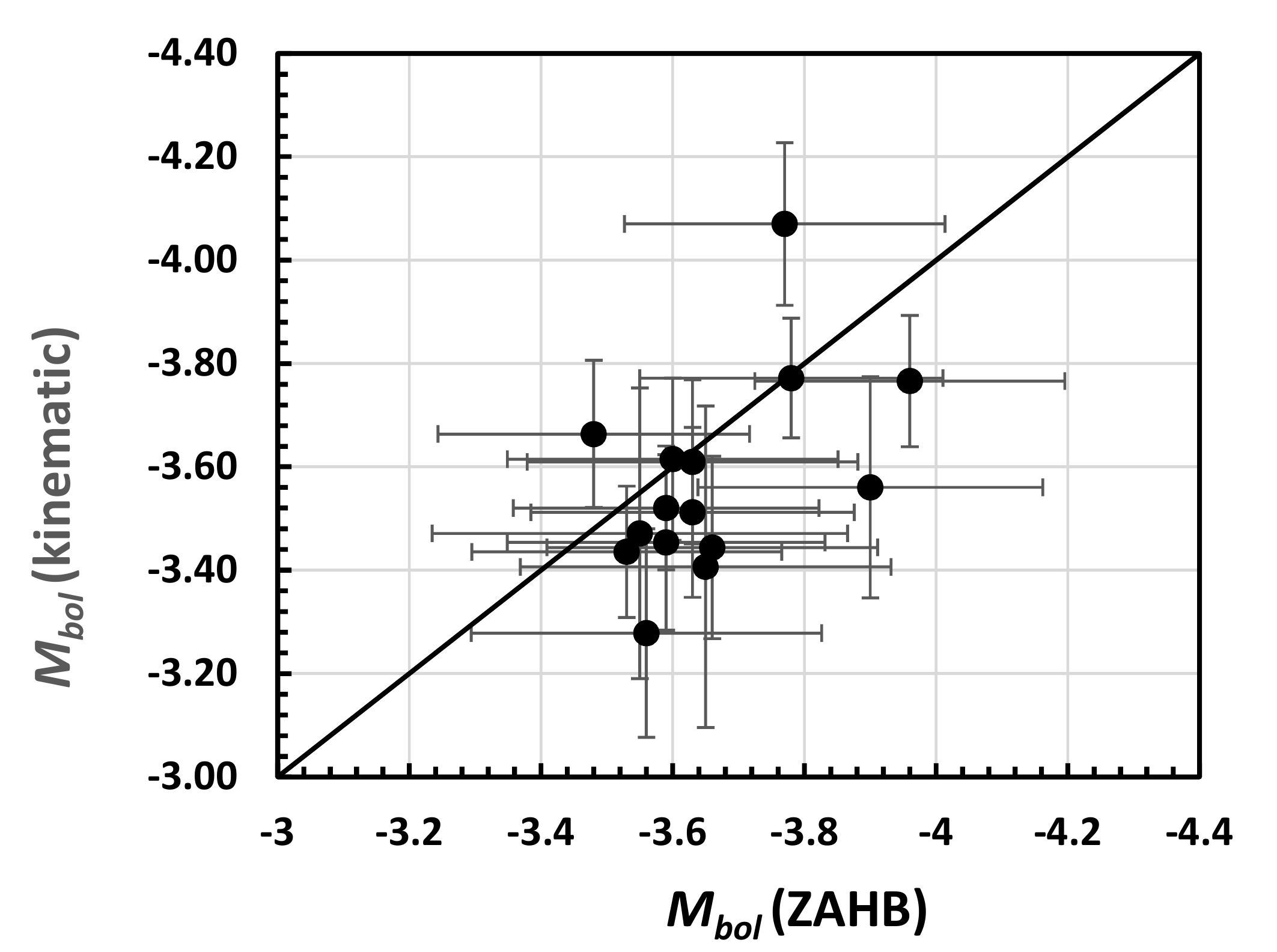}
   \caption{The tip bolometric magnitudes of 16 GGCs obtained by assuming the ZAHB distance scale (horizontal axiis)  are compared to those obtained assuming the kinematic distances \citep[][verical axis]{baumgardt2019}.}
   \label{fig:conf_distance}
\end{figure}

\begin{figure}
   \centering
   \includegraphics[width=9.0cm]{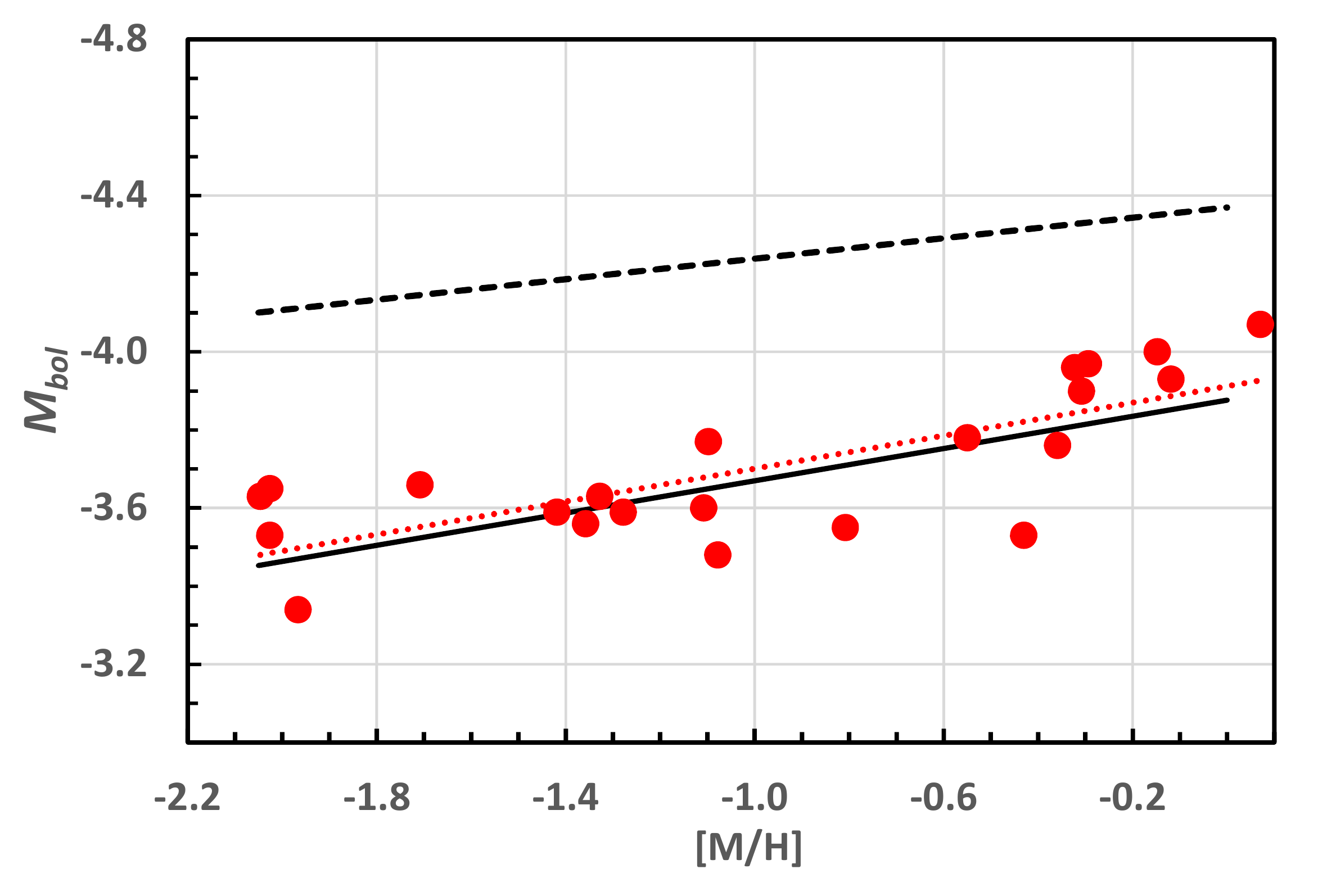}
   \caption{The tip bolometric magnitude derived from optical and IR photometry of 22 GGCs (red circles) are compared to our theoretical predictions for $g_{13}=0$ (black-solid line) and $g_{13}=4$ (black-dashed line). The red-dotted line represents the least square fit of the 22 observed $M_{bol}$. }
   \label{fig:gc_th}
\end{figure}

The RGB tip bolometric magnitude of 22 GGCs, as estimated under different assumptions for the distances and the photometric samples, are listed in columns 6 to 8 of table \ref{tab:clusters}. Here, the $1\sigma$ error includes the uncertainty due to i) the observed apparent magnitude (section \ref{sec:VI} and \ref{sec:IR}), ii) the statistical fluctuations of the magnitudes of the brightest RGB stars (section \ref{sec:tipobs}) and iii) the cluster distance (section \ref{sec:absolutem}). Figure \ref{fig:conf_distance} compare the tip bolometric magnitudes obtained by assuming the distance moduli based on the observed ZAHB with those derived by means of the  kinematic properties. With few exceptions, the differences are generally smaller than the quoted errors. For 47 Tuc, which is one of the two clusters for which we have the most complete sample of bright stars, i.e., VIJHK data of about 200 stars in the brightest 2.5 magnitude of the RGB, the kinematic distance modulus practically coincides with that obtained from the GAIA parallax (and from the ZAHB luminosity), while for NGC 362 it is 0.16 mag larger. For the third best studied cluster, M5, we do not have a reliable estimation of the parallax, but the distance estimated from the ZAHB is in excellent agreement with that obtained by fitting the kinematic properties.            
Observed and predicted tip bolometric magnitudes are compared in figure \ref{fig:gc_th}. The black-solid  line represents our theoretical prediction for $g_{13}=0$ (no-axions), while the dotted line is the linear regression of the 22 data points (squares).  The latter is about 0.04 mag brighter than the theoretical expectation (at $[M/H]=-1$).  This difference could be interpreted as a hint in favor of a small extra-cooling. even if it is of the same order of magnitude of the estimated theoretical error. In any case the axion-electron coupling should be very weak. The theoretical expectation for  $g_{13}=4$ is also shown (black-dashed line). Note that this value of the axion-electron coupling is indicative of the previous upper bound obtained by \citet{viaux2013}. The observed tip bolometric magnitudes clearly exclude such a large coupling. 

In order to estimate the most probable value of  $g_{13}$ and its upper bound, we have calculated, for each clusters in our sample,  the following probability functions:
\begin{equation*}
l_j= \frac{1}{A_j}f_j,
\end{equation*}
\noindent
where: 
\begin{equation*}
f_j = \exp \left [- \frac{(M_{obs}-M_{theo}(g_{13}))^2}{\sigma^2_{obs}+\sigma^2_{theo}+\sigma^2_{age}+\sigma^2_{Y}+\sigma^2_{[M/H]}} \right ],
\end{equation*}
\noindent
and:
\begin{equation*}
A_j = \int_{0}^{\infty}f_j dg_{13},
\end{equation*}
is a normalization factor. 
Here $\sigma_{obs}$ are the errors of the tip absolute magnitude reported in table \ref{tab:clusters}, while $\sigma_{theo}$ represents the cumulative theoretical uncertainty (see \ref{sec:theory}). In addition, we have also considered the uncertainties on the chemical composition, i.e., $Y$ and  $[\rm{M}/\rm{H}]$, and that on the cluster age (see section \ref{sec:absolutem} and table \ref{tab:clusterparam}).
Then, the cumulative likelihood is: 
\begin{equation*}
L(g_{13}) = \frac{1}{A} \exp\left ( -{\bf \Delta^T C^{-1} \Delta} \right ) 
\end{equation*}
\noindent
where $\Delta$ is a vector containing the difference between the observed and the predicted tip bolometric magnitude and $C$ is the error matrix. The latter has been computed following the procedure described in \citet{dagostini1994}. In particular, we have considered the covariance due to the correlations introduced by the calibration of the zero-point of distance scale and bolometric corrections, as well as that due to the theoretical uncertainties (see section \ref{sec:errteo} and \ref{sec:absolutem}). 

\begin{figure}
   \centering
   \includegraphics[width=9.0cm]{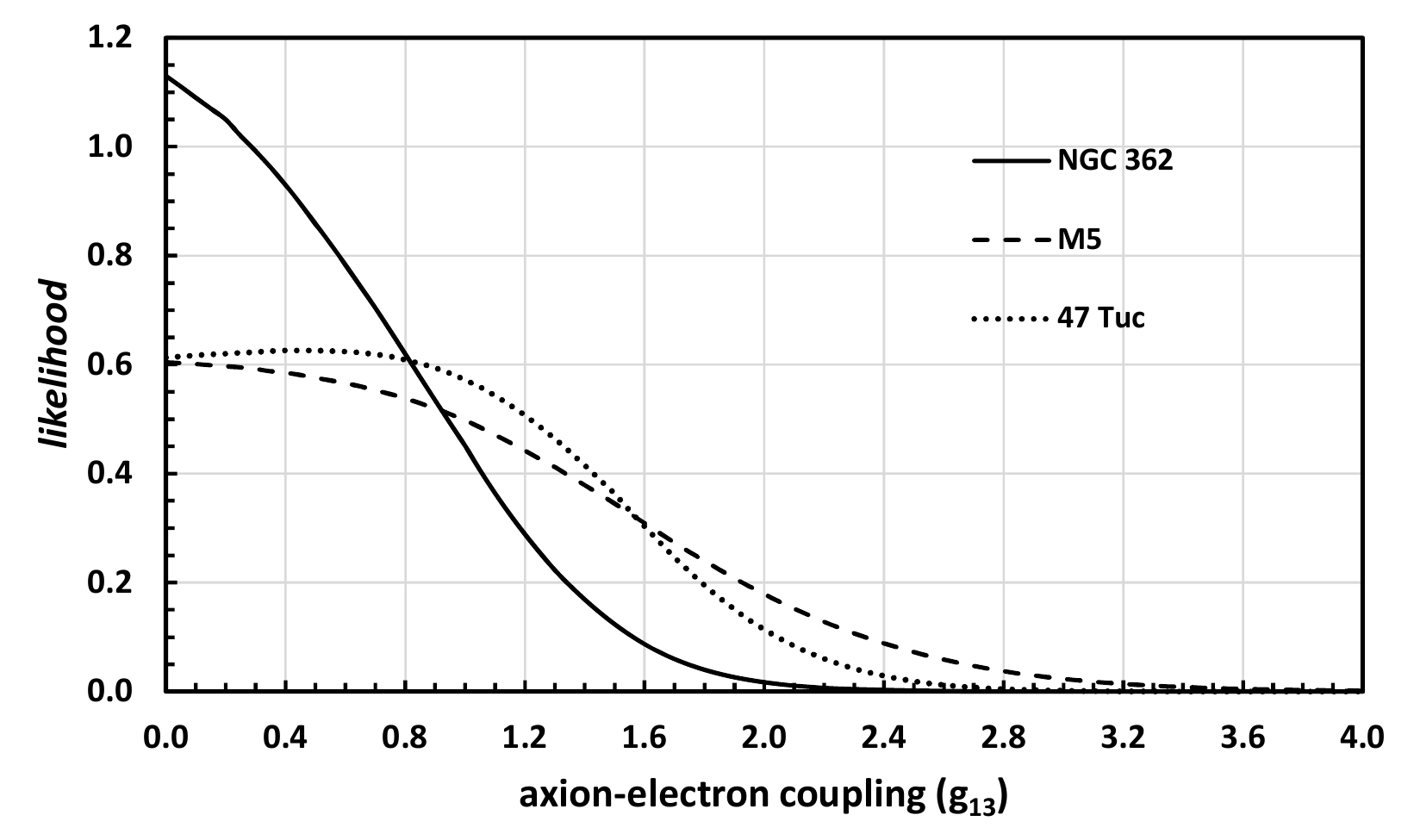}
   \caption{Likelihood functions for M5, 47 Tuc and NGC 362. }
   \label{fig:like_single}
\end{figure}

\begin{figure}
   \centering
   \includegraphics[width=9.0cm]{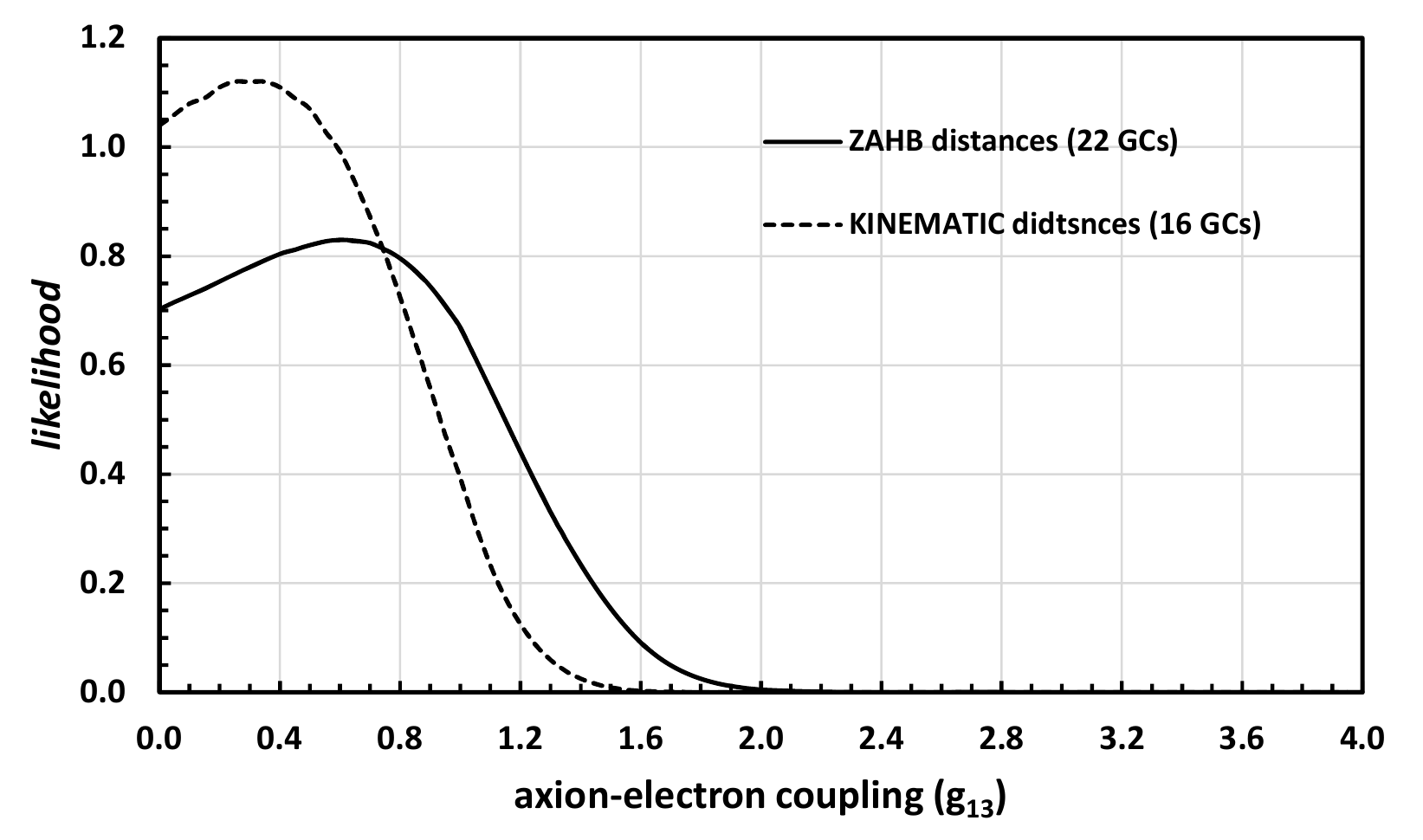}
   \caption{Cumulative likelihood functions, as obtained under different assumptions for the distance scale.}
   \label{fig:like_cum}
\end{figure}

The $l_j$ probability functions for the 3 best studied clusters, i.e.,  M5, 47 Tuc and NGC362, are shown in figure \ref{fig:like_single}. Note that for these 3 clusters both optical and near-IR photometric samples are available. Then, the observed tip bolometric magnitude here adopted is a weighted average of those independently obtained from the VI and the JK samples.   Similarly,  the cumulative likelihood functions, as obtained by assuming the ZAHB distances (22 clusters, solid line) or the kinematic distances (16 clusters, dashed line) are shown in figure \ref{fig:like_cum}.

The best values and the 95\% C.L. upper bounds for the axion-electron coupling ($g_{13}$)  are reported in column 4 and 5 of table \ref{tab:results}. The upper bounds have been obtained by imposing the conditions:
\begin{equation*}
\int_{u.b.}^{\infty} l_j dg_{13} = 0.05  \;\; or \; \int_{u.b.}^{\infty} L dg_{13} = 0.05
\end{equation*}
\noindent
where $u.b.$ is the $g_{13}$ upper bound. In practice, values of $g_{13}$ greater than u.b. are excluded with 95\% of confidence. 

  The two distance scales lead to a similar conclusion, namely: there is an hint for a weak axion-electron coupling, $g_{13}\sim 0.05$, with a stringent upper bound of $g_{13}<1.5$ (95\% confidence). 

\begin{table*}
\centering
\caption{Results of the likelihood analysis. The most probable values of the $g_{13}$ parameter are listeded in column 4, while the corresponding upper bounds (95\% confidence) are in column 5. The first 3 rows report the results obtained for the 3 best studied clusters, those for which we have used both optical and near-IR catalogs, while the last 2 rows list the results of the cumulative analysis. Different distance scales have  been used. In particular: ``ZAHB'' stays for distances based on the observed ZAHB luminosity, 2) ``parallax'' indicates distances based on Gaia DR2 parallaxes \citep{chen2018} and 3) ``kinematic'' indicates distances based on the best fit of the Gaia DR2 kinematic properties \citep{baumgardt2019}.}
\label{tab:results}
\begin{tabular}{lcccc}
\hline
     & photometric  & distance  & $g_{13}$ & $g_{13}$  \\
     & sample       & scale  & best value & bound  \\
\hline 
    M5     & VI+JK  & ZAHB     & 0    & 2.30 \\
    47 Tuc & VI+ JK & parallax & 0.45 & 1.87  \\
    NGC 362& VI+JK  & parallax & 0    & 1.37  \\
    22 GGCs & JK     & ZAHB     & 0.60 & 1.48 \\
    16 GGCs & JK     & KINEMATIC & 0.35 & 1.15 \\
\hline    
\end{tabular}
\end{table*}


\section{Conclusions}\label{sec:conclusions}
Basing on VIJK photometries of bright RGB stars and state-of-the-art distance and metallicity scales, we have determined the absolute magnitude of the RGB tip for a sample of 22 GGCs. An accurate evaluation of all the uncertainties has been performed. We have also revised the corresponding model predictions, by analyzing  the present theoretical and experimental knowledge of the relevant input physics, such as nuclear reactions, neutrino emission rates, thermodynamic properties of stellar plasma and the like. These theoretical predictions are, in general, in good agreement with the observed tip bolometric magnitudes, even if the latter are $\sim 0.04$ mag brighter, on the average. This small shift between theoretical and observed tip bolometric magnitude could be the consequence of a weak axion-electron coupling. Following this hypothesis, we have computed additional stellar models by including the production of thermal axions. Note that due to the high electron degeneracy, Bremsstrahlung is the most relevant thermal process capable to produce a sizeable flux of axions from the core of RGB stars. Therefore, by means of a cumulative likelihood analysis, as obtained considering all the 22 clusters of our sample, we have estimated the coupling parameter that determines the strength of the energy sink induced by the production of Bremsstrahlung axion. We find that the likelihood probability is maximized for 
$g_{ae}\sim 0.60^{+0.32}_{-0.58}\times 10^{-13}$. Moreover, we find 
an upper bound for this parameter of $g_{ae}=1.48\times 10^{-13}$, with 95\% confidence.   

This new bound represents the more stringent constraint for the axion-electron coupling available so far. Indeed, in addition to the work of \citet{viaux2013}, who report $g_{ae}<4.3\times 10^{-13}$ from the luminosity of the RGB tip of the globular cluster M5, other astrophysical constraints we find in the extant literature are those of   
\citet{MB2014} \citep[see also][]{isern2018}, reporting $g_{ae}<2.8\times 10^{-13}$, as obtained from an analysis of the  white dwarfs (WDs) luminosity functions, and \citet{corsico2016}, who find $g_{ae}<7\times 10^{-13}$, from the period drift of pulsating white dwarfs. 

Concerning the direct search for axions, the recent years have seen an impressive experimental efforts \cite[for a review of the axion experimental landscape see][]{irastorza2018,DiLuzio:2020wdo,Sikivie:2020zpn}. However, the axion coupling to electrons is particularly difficult to probe experimentally.
The most stringent upper bounds have been obtained by the XENON100 collaboration \citep{xenon}, $g_{ae} < 7.7\times 10^{-12}$ (90 \% CL), LUX~\citep{Akerib:2017uem}, $g_{ae} < 3.5\times 10^{-12}$, and 
PandaX-II~\citep{Fu:2017lfc}, $ g_{ae}<4\times 10^{-12} $.
These bounds are not yet competitive with the stellar bounds on this coupling. 

Before submitting the present paper, there was an announcement from the XENON1T collaboration of a possible detection of solar axions \citep{aprile2020}. For values of the axion-photon coupling not excluded by previous experiments \citep[e.g.,][]{cast2017}, 
they find $27<g_{ae}/10^{13}<37$ (90 \% confidence), 
a value at least 20 times higher than the upper bound we obtained 
in this work \citep[see also][]{DiLuzio:2020jjp}.
Alternative interpretations of the signal detected by XENON1T have been already proposed \citep[see the discussion in][]{aprile2020},
among which the possibility that this signal is due to a dark matter ALP with mass of a few keV and  a  coupling  to  electron $g_{ae}\sim 10^{-13}$ \citep{takahashi2020}. In that case, these ALPs may constitute  all or  some  fraction  of the  local dark  matter.  Such an explanation would remove the tension of the XENON1T detection with our work.       

Let us finally remark that the present upper bound for the axion-electron coupling remains valid also in case of additional energy sinks active during the RGB evolutionary phase, not included in the present models, such as a non-zero neutrino magnetic moment or a sizeable axion coupling with photons. On the contrary, the existence of these additional energy sinks would affect the hint on $g_{ae}$. For example, in section \ref{sec:axions} we have recalled that at the high density of the core of a RGB star the Primakoff process is suppressed. Nevertheless its effect on the RGB tip luminosity may be not completely negligible. In the most extreme case of an axion-photon coupling of $g_{a\gamma}=6\times 10^{-11}$ GeV$^{-1}$, i.e., the upper bound independently obtained from the solar-axions by the CAST collaboration \citep{cast2017} and from the horizontal branch lifetimes \citep{ayala2014}, the resulting upward shift of the RGB tip would explain most of the small difference between the standard theoretical predictions (no-axions) and the last square fit of the observed tip luminosity (the solid and the dotted curves in figure \ref{fig:gc_th}).  In this framework, we cannot exclude a much smaller, at least 0, axion-electron coupling. 

 
\begin{acknowledgements}
We are in debt with F. Capozzi and G. Raffelt for a careful reading of our manuscript. We acknowledge that after our submission to A\&A, they uploaded on the ArXiv an independent study where they reached conclusions similar to ours. 
Our work has been supported by the Agenzia Spaziale Italiana (ASI) and the Instituto Nazionale di Astrofsica (INAF) under the agree-ment  n.   2017-14-H.0  - attivit\`a  di  studio  per  la  comunit\`a  scientifica  di  Astrofisica  delle  Alte Energie  e  Fisica Astroparticellare. I.D. research  is also supported by the MICINN-FEDER project  PGC2018-095317-B-C21, while 
A.M. is partially supported by
the Istituto Nazionale di Fisica Nucleare (INFN),
through the ``Theoretical Astroparticle Physics'' project,
and by the research grant number 2017W4HA7S:
``Neutrino and Astroparticle Theory Network'', under
the program PRIN 2017 funded by the Italian Ministero
dell'Universit\'a e della Ricerca (MUR).

\end{acknowledgements}

\bibliographystyle{aa}
\bibliography{axions.bib} 


\end{document}